\documentclass[12pt]{article}\usepackage[hyperfootnotes=false]{hyperref}
\usepackage{epsfig}
\usepackage{float}
\usepackage{dsfont}
\usepackage{bbold}

 \usepackage{dsfont}

\usepackage{caption}
\usepackage{subcaption}

\usepackage{amsmath}
\usepackage{amssymb}
\usepackage{graphicx}
\newlength{\abstractwidth}
\renewcommand{\thefootnote}{\fnsymbol{footnote}}
\renewcommand{\thanks}[1]{\footnote{#1}} 
\newcommand{\starttext}{
\setcounter{footnote}{0}
\renewcommand{\thefootnote}{\arabic{footnote}}}

\newcommand{\be}{\begin{equation}}
\newcommand{\bea}{\begin{eqnarray}}
\newcommand{\eea}{\end{eqnarray}}
\newcommand{\beq}{\begin{equation}}
\newcommand{\ee}{\end{equation}}

\DeclareMathOperator{\Tr}{Tr}
\def\eq{&=&}

\def\la{\langle}
\def\ra{\rangle}

\def\simleq{\; \raise0.3ex\hbox{$<$\kern-0.75em
\raise-1.1ex\hbox{$\sim$}}\; }
\def\simgeq{\; \raise0.3ex\hbox{$>$\kern-0.75em
\raise-1.1ex\hbox{$\sim$}}\; }

\def\bi{\begin{itemize}}
\def\ei{\end{itemize}}

\def\sc{\setcounter{equation}{0}}

\def\CA{{\cal{A}}}

\def\CC{{\cal{C}}}

\def\CI{{\cal{I}}}

\def\CQ{{\cal{Q}}}
\def\CR{{\cal{R}}}

\def\CA{{\cal{A}}}

\def\kl{$k$-local}
\def\sa{$\CA$}

\def\bn{\bigskip \noindent}

\def\suk{SU(2^K)}
\def\sk{$SU(2^K)$}

  \def\Ud{U^{\dag}}
   \def\dU{\dot{U}}
    
    \def\dUd{\dot{U}^{\dag}}

\makeatletter
\g@addto@macro\normalsize{%
  \setlength\abovedisplayskip{10pt}
  \setlength\belowdisplayskip{20pt}
  \setlength\abovedisplayshortskip{10pt}
  \setlength\belowdisplayshortskip{20pt}
}
\makeatother

\usepackage{color}

\renewcommand{\title}[1]{\vbox{\center\LARGE{#1}}\vspace{5mm}}
\renewcommand{\author}[1]{\vbox{\center#1}\vspace{5mm}}
\newcommand{\address}[1]{\vbox{\center\em#1}}

\begin{document}
  
\begin{titlepage}

\rightline{}
\bigskip
\bigskip\bigskip\bigskip\bigskip
\bigskip

\centerline{\Large \bf {The Second Law of Quantum Complexity}}

\bn

\bigskip

\bigskip
\begin{center}

\author{Adam R. Brown and Leonard Susskind}

\address{Stanford Institute for Theoretical Physics and Department of Physics, \\
Stanford University, Stanford, CA 94305-4060, USA}

\end{center}

\begin{center}
\bf     \rm

\bigskip

\end{center}

\begin{abstract}
We give arguments for the existence of a thermodynamics of quantum complexity that includes a ``Second Law of Complexity". To
guide us, we derive a correspondence between the computational (circuit) complexity
of a quantum system of $K$ qubits, and the positional entropy of a related classical system with $2^K$ degrees of freedom.  We also argue that the kinetic entropy
of the classical system is equivalent to the Kolmogorov complexity of the quantum
Hamiltonian. We observe that the expected pattern of growth of the complexity of the quantum system parallels the growth of entropy of the classical system. We argue
that the property of having less-than-maximal complexity (uncomplexity) is a resource that can be
expended to perform directed quantum computation.

Although this paper is not primarily about black holes, we find a surprising  interpretation of the uncomplexity-resource as the accessible volume of spacetime behind a black hole horizon.

\medskip
\noindent
\end{abstract}

\let\thefootnote\relax\footnotetext{emails: \tt{adambro@stanford.edu, susskind@stanford.edu}}

\end{titlepage}

\starttext \baselineskip=17.63pt \setcounter{footnote}{0}

\vfill\eject

\tableofcontents

\vfill\eject

\sc
\section{Quantum Complexity and Classical Entropy} \label{Sec: QC and CE}

Complexity theory, particularly its quantum version, is a new and relatively unknown mathematical  subject to most 
physicists\footnote{Including the authors of this paper.}.
 It's a difficult subject with few quantitative results and, at least for the moment, no experimental guidance. Our original interest in complexity began with questions about black holes 
\cite{Susskind:2014rva,Stanford:2014jda,Susskind:2014jwa}, but broadened into the issue of 
what happens to quantum systems between the time they reach maximum entropy and the much later time they reach maximum complexity.

The mainstream goals of complexity theory are to organize tasks into broad qualitative complexity classes. Our main focus will be somewhat different. Our concern is with the quantitative behavior of complexity as a system evolves. The two types of questions are by no means unconnected but they are different and probably require different tools.
In this paper we will consider whether physics---especially statistical mechanics and thermodynamics---may be useful  for analyzing  the growth and evolution of complexity in generic quantum systems.

In particular we are interested in whether there is an analog, involving quantum complexity, for the second law of thermodynamics. In \cite{Brown:2016wib,Susskind:2015toa}
such a Second Law of Complexity was conjectured, and  invoked for the purposes of diagnosing the transparency of horizons, i.e., the absence of firewalls \cite{Almheiri:2012rt}. It was argued \cite{Susskind:2015toa} that opaque horizons with firewalls are associated with  states of decreasing complexity, and that as long as the complexity of the quantum state   increases, the  horizon will be transparent. A Second Law of Complexity would ensure that a black hole formed from natural processes will have increasing complexity and therefore a transparent horizon, at least for an exponentially long time.

In this paper we argue that the Second Law of Complexity for a quantum system $\CQ$ is a consequence of the second law of thermodynamics for an auxiliary classical system $\CA$.

Two distinct notions of quantum complexity will be discussed. The first, denoted $\CC$, is computational complexity, also called circuit complexity or gate complexity. It measures the minimum number of gates required to prepare a given unitary operator or a given state\footnote{Throughout this paper the term `state' will always mean a pure state. More general density matrices will be called mixed states.} from an unentangled product state. 
The second is Kolmogorov complexity, denoted $\CC_{\kappa},$ whose relevance will become clear in Sec.~\ref{KC_and_KE}.

\subsection{The Evolution of Complexity}

\begin{figure}[H]
\begin{center}
\includegraphics[width=5in]{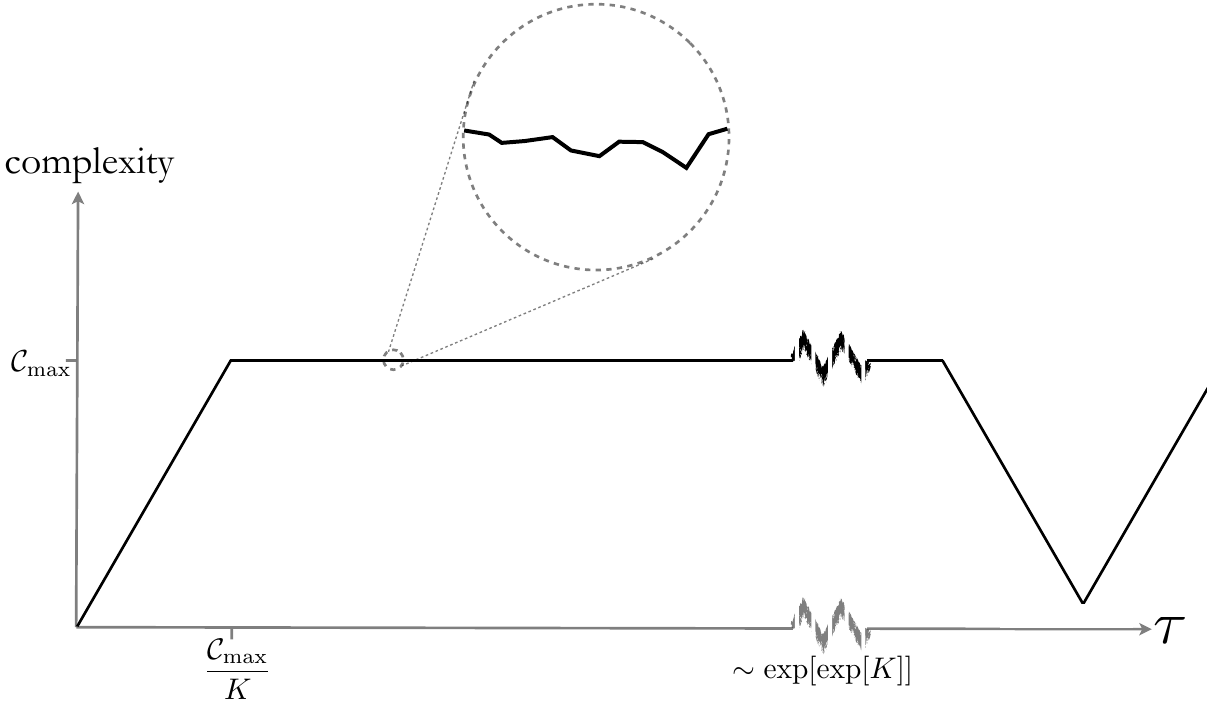}
\caption{The conjectured evolution of quantum complexity of the operator $e^{-iHt}$, where $H$ is a generic time-independent $k$-local Hamiltonian. The complexity increases with rate $K$, and then saturates at a value exponential in $K$. It fluctuates around this value. Quantum recurrences occur on a timescale that is double-exponential in $K$; a very rare, very large fluctuation brings the complexity down to near zero. This figure would also describe the entropy of a classical chaotic system with $\exp[K]$ degrees of freedom.}
\label{f1}
\end{center}
\end{figure}

The object of interest is the time-development operator $U(t) = e^{-iHt}$, for a generic $k$-local system of the type that model black holes. 
The question of interest is how the computational complexity\footnote{The concept of computational complexity that we are using is essentially the same as quantum circuit complexity, i.e., the minimal number of quantum gates needed to prepare a given unitary operator.}  of $U(t)$ evolves with time.
Both black hole and quantum circuit  considerations  suggest the following conjecture summarized in Fig.~\ref{f1}.

\bn
The complexity $\CC(t)$ grows linearly as
\be
\CC(t)= K t
\label{Kt/2}
\ee
for a time exponential in $K.$ At $ t\sim e^K$ the complexity reaches its maximum possible value $\CC_\textrm{max}$ and flattens out for a very long time. This is the period of \it complexity equilibrium \rm   \cite{Susskind:2015toa} during which the complexity fluctuates about the maximum,
\be 
\CC_\textrm{max} \sim e^K. \label{eq:Cmax}
\ee
On a much longer timescale of order $\exp[e^K]$ quantum recurrences quasiperiodically return the complexity to sub-exponential values. All of this is a conjecture which at the moment cannot be proved, but which can be related to other complexity conjectures  \cite{Aaronson}.

The pattern  described above is  reminiscent of the evolution of classical entropy. Starting a classical system in a configuration of low entropy (all the molecules in the corner of the room)
 the subsequent evolution, as the gas comes to equilibrium, follows a similar curve to Fig.~\ref{f1}, but for entropy---not complexity. However, for the classical case the linear growth of  entropy will persist for only a  time polynomial (in the number of degrees of freedom), the maximum entropy will also be of order the number of degrees of freedom, and the recurrence time will be simply exponential and not doubly exponential.

A simple and concise way to express the parallel is: 

\bn

\it The quantum complexity for a system of $K$ qubits behaves in a manner similar  to the entropy of a classical system with $2^K$  degrees of freedom.

\rm

\bn
The primary goal of this paper  is to understand this similarity.\\

In \cite{Brown:2016wib} a two-dimensional toy analog model for complexity was conjectured. (We recommend that the reader first read \cite{Brown:2016wib} before this paper.) The motivation for the toy model was Nielsen's geometric approach to complexity. Here we are going to consider the far more complex case  based directly on a version of Nielsen's high-dimensional geometry \cite{Nielsen}\cite{Dowling}.  \\

Another goal that we discuss is the construction of a resource theory of complexity in which the relevant thermodynamic resource would be the gap between the complexity of a system and the maximum possible complexity---the `uncomplexity'. This is expended by performing directed quantum computation, which means reducing the relative complexity of the initial state and the target state. We suggest that this resource can, under appropriate conditions, be used to do directed  quantum computation in much the same way that in conventional thermodynamics free energy is used to do directed work. 
We will have more to say about this in Sec.~\ref{Sec: combining}. \\

\noindent A guide to the notations and conventions and units used in this paper can be found in Appendix~\ref{sec:terminology}.

\sc
\section{The Quantum System $\CQ$} \label{sec: Qsystem}
\subsection{Randomness}\label{sec: randomness}

There are many problems in both classical and quantum physics that are extremely difficult when particular instances of the problem are considered. The strategy of averaging over ensembles of instances sometimes allows conclusions to be drawn about generic behavior that would not be possible for specific cases. A particular example, which has generated recent interest, is the SYK approach to scrambling. By averaging over an appropriate ensemble of time-independent Hamiltonians it is possible to show that almost all  such Hamiltonians saturate the fast-scrambling bound \cite{Sekino:2008he}\cite{Maldacena:2015waa}. Potentially this kind of averaging can also  be applied  to questions about the evolution of complexity. 

Another type of randomness is stochastic randomness in which  a time-dependent statistically fluctuating (noisy)  Hamiltonian is averaged over.
Generally the more one averages over the easier it is to draw conclusions, and indeed stochastic averaging is easier than averaging over time-independent Hamiltonians. Of course, if our interest is in the behavior of  time-independent Hamiltonians (as it is in this paper) it is not entirely clear that the lessons we learn from stochastic behavior are applicable.

\subsection{$k$-locality}

The systems we will consider are constructed from qubits and have a type of dynamics called \kl. Other than that they are very generic. The building blocks of a \kl \ Hamiltonian are Hermitian operators that involve at most $k$ qubits. The term ``weight" applied to an operator means the number of single qubit  factors that appear in the operator. A \kl \ Hamiltonian is one that contains terms of no higher weight than $k$. 
  Ordinary lattice Hamiltonians with nearest neighbor couplings are $k$ local---in fact they are 2-local. But $k$-locality does not assume any kind of spatial locality. For example we may have Hamiltonians in which any pair of qubits directly interact. The general form of an exactly \kl \ Hamiltonian built out of standard qubits is\footnote{By `exactly $k$-local' we mean a Hamiltonian that is a sum of terms each of which acts on exactly $k$ qubits (not that each term acts on $k$ \emph{or fewer} qubits). See clarification 1 in Appendix~\ref{Sec:clarifications}.}

\be 
H = \sum_{i_1< i_2<...<i_k} \sum_{a_1 = \{x,y,z\}} \ldots \sum_{a_k = \{x,y,z\}}
J_{i_1, i_2,...,i_k}^{a_1, a_2,...,a_k} \
\sigma_{i_1}^{a_1}  \, \sigma_{i_2}^{a_2}.... \,\sigma_{i_k}^{a_k}.
\label{eq:$k$-local}
\ee

The SYK model is another \cite{Sachdev:1992fk,Kitaev,Polchinski,Maldacena:2016hyu} type of \kl \ system which is built out of real anti-commuting degrees of freedom $\chi_i,$
\be 
\{ \chi_i , \chi_j \} =  \delta_{ij}.
\ee
The SYK Hamiltonian is similar to Eq.~\ref{eq:$k$-local},

\be 
H = \sum_{i_1< i_2<...<i_k}
J_{i_1, i_2,...,i_k} \
\chi_{i_1} \ \chi_{i_2}....\chi_{i_k}.
\label{eq:$Fk$-local}
\ee 
 
The SYK model is \kl \ when written in terms of fermions, but if we try to rewrite it in terms of standard qubit operators it will be highly non-local. Despite this,  most of what we  describe applies to it. For definiteness we will illustrate the principles for systems of standard qubits with Hamiltonians of the form Eq.~\ref{eq:$k$-local}.
 
There is however a caveat. The SYK model is usually studied at low temperature where it has an approximate conformal invariance and behaves roughly like a near-extremal charged black hole. At low temperature, standard qubit models are different; for example they may have spin-glass behavior. Our interest will instead be in the high-temperature behavior where we expect both kinds of models behave somewhat similarly to uncharged Schwarzschild black holes.  At high temperature the conformal invariance does not play a role.
 
 It should be noted that for systems of fermions or qubits the high-temperature limit is not a high-energy limit.   The energy and entropy per qubit do not go to infinity at infinite temperature; in fact the entropy per degree of freedom does not change much between the usual SYK low-temperature regime and infinite temperature. It is also true that the ratio of the Lyapunov exponent to the
 energy-per-qubit tends to a finite constant as temperature increases. This is in contrast to the ratio of the Lyapunov exponent to temperature, which goes to zero in the high-temperature limit. This means that at higher temperatures the Maldacena-Shenker-Stanford bound
\cite{Maldacena:2015waa} is not tight and a stronger bound might be expressed in terms of the energy-per-qubit rather than the temperature. 

Hamiltonians of the type Eq.~\ref{eq:$k$-local} are very common. They include lattice systems, for which the couplings are non-zero only for nearest neighbors on some ordinary lattice. But these ``condensed matter'' Hamiltonians are very rare in the space of the couplings. The generic \kl \ Hamiltonian  is a fast-scrambler \cite{Hayden:2007cs}\cite{Sekino:2008he}, meaning that every qubit is coupled to every other qubit, but only through terms involving at most $k$ qubits. We will be interested in this generic case. Averages over the $J$'s  will be dominated by  fast scramblers.

For now  assume that the $J$'s are known definite numbers, but keep in mind that  the trick of averaging over Hamiltonians may make otherwise impossible problems tractable. 
To simplify the notation we will write Eq.~\ref{eq:$k$-local} in the schematic form
\be 
H = \sum_{I} J_I \sigma_I,
\label{H=Jsig}
\ee
where $I$ runs over all $(4^K-1)$ generalized  Pauli 
operators\footnote{By the generalized Pauli operators we mean the set of $3K$ Pauli operators $\sigma_i^a$ together with all possible  products, with no locality restrictions.} with the proviso that only the \kl \ couplings are non-zero.

\subsection{The Quantum System}

The quantum system $\CQ$ consists of $K$ qubits interacting through a \kl \
Hamiltonian of the form Eq.~\ref{eq:$k$-local}. Adapting the discussion to fermionic degrees of freedom, as in Eq.~\ref{eq:$Fk$-local}, should be straightforward.

 We will not be interested in any particular Hamiltonian; following Sachdev-Ye and Kitaev we will consider the  properties of the system when averaged over a Gaussian statistical ensemble of the $J$-coefficients. The probability for the \kl \ couplings $J_I$ to take a specified set of values is,
\be 
P(J) = \frac{1}{Z} e^{-\frac{1}{2} B_a \sum_I J_I^2}.
\label{P(J)}
\ee
The constant $B_a$ determines the variance of the distribution. The non-\kl \ couplings are assumed to be zero. 

The space of states is $2^K$ dimensional. Unimodular unitary operators are represented by
$2^K\times 2^K$ matrices in \sk. These matrices can be thought of in two ways. The first is as operators acting on the state space of the $K$ qubits. The second is as wavefunctions of maximally entangled states of $2K$ qubits.
In this latter sense the identity matrix represents a thermofield-double (TFD) state at infinite temperature. As such it is dominated by the highest energy states of the system. 

Let us consider the variance of the Hamiltonian  in the infinite temperature TFD state,
\bea 
(\Delta H)^2 \eq \Tr H^2 \cr \cr
\eq \Tr \sum_I \sum_J \sigma_I \sigma_J J_I J_J.
\eea
(Here and throughout this paper we normalize $\Tr$ so that $\Tr \mathds{1} = 1$.)
The generalized Pauli matrices $\sigma_I$ satisfy
\be 
\Tr \sigma_I \sigma_J = \delta_{IJ}. 
\ee
Thus 
\be 
(\Delta H)^2 = \sum_I J_I^2.
\label{E2=J2}
\ee
Note that the average of the Hamiltonian itself is zero since all the terms in $H$ have zero trace. We will use the notation $E$ to represent the  energy relative to the ground state. This is not zero. The variance of $E$ is the same as the variance of $H.$

The normalization of $J$ is a convention related to the normalization of time. We choose it by observing that 
fast scramblers are models for neutral static  black holes. It is a general fact about such black holes that their dimensionless Rindler energy  $E$ (defined relative to the ground state), and  the variance of the dimensionless energy $(\Delta E)^2$  are both equal to the entropy. For the infinite temperature TFD state the entropy of each copy is $S=K.$ It follows that the distribution of $J$'s should satisfy
\be
\sum J_I^2\Bigl|_\textrm{av} =  K  = E,
\label{J2=K}
\ee 
where the average in $\Bigl|_\textrm{av}$ is an ensemble average.

\bn

\bn

If the Hamiltonian is exactly \kl \  then the number of $J$-coefficients is
\be 
N_J = 3^k {K \choose k}  \approx \frac{(3K)^k}{k!}.
\label{miss3}
\ee
Letting $J^2$ be the variance of any of the $J_I$, Eq.~\ref{J2=K} gives
\be 
J^2 = \frac{k!}{3^k K^{k-1}}.
\label{syk variance}
\ee

The same argument, when applied to the SYK model, correctly  gives the variance, the only difference being the absence of the factor $3^{-k}$ in the SYK case\footnote{The factors of $3$ are due to there being three  Pauli matrices.}. 
 The relevance of these facts will become clear when we study the thermodynamics of the auxiliary system $\CA.$

Hamiltonians of the form Eq.~\ref{H=Jsig} can easily be generalized to stochastic evolution by allowing the $J$'s to have a time-dependence governed by a stochastic probability distribution. The resulting ``Brownian circuits" were discussed in \cite{Lashkari:2011yi}.

\sc
\section{The  Classical  System  $\CA$}\label{sec: classical system}

For the moment we will ignore issues of complexity and define a classical system  that represents the evolution of a quantum system as the motion of a non-relativistic particle moving on  \sk. Later we will modify the geometry to a ``complexity geometry" along the lines of \cite{Dowling}.

The space \sk \ is a homogeneous group space generated by $(4^K-1)$ generators which in the Pauli basis are the generalized Pauli operators $\sigma_I.$ Each point on \sk \ corresponds to an element of \sk: it is a particular $2^K$ by $2^K$  unimodular matrix $U.$ Up to an overall constant factor, the unique bi-invariant metric is given by\footnote{Throughout this paper the notation $\Tr$ refers to normalized trace such that $\Tr {\mathds{1}} = 1.$},
\begin{equation} 
dl^2 = \Tr d\Ud dU.
\label{bimetric}
\end{equation}
This metric is called `bi-invariant' since it is invariant under both left- and right-multiplication:  for any $W \in$  \sk \  and $V \in$  \sk, transforming $U$ by 
\be
U\to W^{\dag}UV
\ee
does not change the metric distance in Eq.~\ref{bimetric}.

\subsection{Equations of Motion} \label{subsec:equationsofmotion}

The time evolution of the system with Hamiltonian Eq.~\ref{H=Jsig} defines a moving point $U(t)$ which we may think of as the motion of a fictitious particle moving on \sk.
The particle starts at the point $U= \mathds{1}$, i.e., the identity matrix. The motion can be represented by ordinary classical mechanics. Begin with the Schrodinger equation for $U(t),$
\be 
i\dU = H U. 
\label{schrodinger}
\ee
This is a first order (in time) equation and, given the Hamiltonian, through every point $U$ there is a unique trajectory. 
We would like to write this in a way that does not make reference to a specific Hamiltonian. To that end we first
solve for $H,$
\be
H = i\dU \Ud.
\label{Hsolve}
\ee
Differentiating Eq.~\ref{schrodinger}  with respect to time and then plugging in Eq.~\ref{Hsolve} gives the equation of motion
\be 
\ddot{U}-\dU \Ud \dU =0.
\label{eqmo}
\ee
This is the second order equation of motion of a non-relativistic particle moving on \sk. It is well known that such motion is along geodesics with constant velocity.
 In terms of general coordinates the equation of motion has the familiar form 
\be 
{\ddot{X}}^M = -\Gamma^M_{AB}\dot{X^A}\dot{X^B},
\label{xquation}
\ee
where $\Gamma^M_{AB}$ are the Christoffel symbols derived from the standard metric on \sk.\\

\noindent In summary, there are two ways to specify the evolution of a unitary under a time-independent Hamiltonian. One is to specify $U(t=0)$ and $H$ and use the first-order equation of motion Eq.~\ref{schrodinger}. The second is to specify $U(t=0)$ and $\dot{U}(t=0)$ and use the second-order equation of motion Eq.~\ref{eqmo}. These two formulations are equivalent, since we can translate between $H$ and $\dot{U}$ using Eq.~\ref{Hsolve}. In what follows we will find it more convenient to work with the second-order equation of motion, which makes no explicit reference to $H$ and instead stores that information in the initial condition $\dot{U}$.

\subsection{Velocity-Coupling Correspondence}\label{sec: VC}
Note that the equation of motion Eq.~\ref{eqmo} no longer makes reference to the Hamiltonian. That information is now encoded in the initial conditions. 
To see how this works
we write the Hamiltonian Eq.~\ref{eq:$k$-local} as
\beq
H = \sum_I J_{I}\sigma_I.
\ee
The Schrodinger equation takes the form
\beq
\dot{U} = - i\sum_I J_{I} \sigma_I \ U.
\label{eq-of-mot}
\ee
We can easily solve for $J_I$,
\be 
J_I = i\Tr \sigma_I \dot{U} U^{\dag}.
\label{Gen=sUU}
\ee
At the origin $U=  \mathds{1}$ we may write
\beq
i\Tr \dot{U}\sigma_I = J_I.
\label{J=sU}
\ee
The left side of this equation is the projection of the initial velocity onto the tangent space axes  oriented along the Pauli basis.  In other words the $J_I$ are the initial values of the velocity components $V_I,$
\be 
J_I = V_I \Bigl|_\textrm{initial}.
\label{VJ-correspondence}
\ee
 We'll  call Eq.~\ref{VJ-correspondence}    Velocity-Coupling correspondence,  or just  V/J-correspondence.
 
A point to emphasize is that the classical mechanics described by the equation of motion Eq.~\ref{eq-of-mot} is not the theory of any particular Hamiltonian. It is the theory of all Hamiltonians of the form Eq.~\ref{eq:$k$-local} with the $J$'s playing the role of initial velocities.

\subsection{Action}
The equations of motion  Eq.~\ref{xquation}, or equivalently Eq.~\ref{eqmo}, may be derived from an action\footnote{The subscript $_a$ refers to the auxiliary system; for a guide to conventions see Appendix~\ref{sec:terminology}.},
\bea 
A_a \eq \int L_a dt \cr \cr
L_a \eq \frac{1}{2} G_{MN}\dot{X}^M\dot{X}^N.
\label{lagrangian}
\eea
In terms of $U$, this action has the simple form
\be
L_a = \frac{1}{2} \Tr \dUd \dU.
\ee

 \subsection{Conservation Laws}
 The  $\CA$-system has a conserved Hamiltonian which is not to be identified with the Hamiltonian of the $\CQ$-system (namely Eq.~\ref{eq:$k$-local}). From the form of the Lagrangian one finds that the auxiliary energy is the same as the Lagrangian,
 \be
 E_a = L_a = \frac{1}{2} \Tr \dUd \dU.
 \ee
 The energy is of course just the familiar non-relativistic expression for a particle of unit mass,
 \be
 E_a = \frac{1}{2}V_a^2.
 \label{E-sub-a}
 \ee
 The other conservation laws follow from the bi-invariance of the  metric. Invariance under right multiplication gives rise to conservation of the matrix elements of the right charges,
 \be 
 Q^R_I = i\Tr \sigma_I \Ud \dU.
 \ee
The left charges,
  \be 
 Q^L_I = i\Tr  \sigma_I  \dU \Ud,
 \ee
  are also conserved,
 but they are not functionally independent of the right charges---the $Q^L_I$s can be written in terms of the $Q^R_I$s and the matrices $U$.

 \subsection{Ergodicity}\label{sec:ergodicity}
 Naively we might expect the motion generated by a generic time-independent $k$-local Hamiltonian to be ergodic on \sk. But in fact the motion is very far from ergodic. To see this, consider writing $e^{- i H t}$ in the energy basis
 \be
e^{- i H t} = \sum_n  e^{-iE_n t} |n \ra \la n| .
 \label{toro}
 \ee
 For a given Hamiltonian there are $2^K$ energy eigenvalues and it follows that
 $U$ moves on a torus of dimension $2^K.$ This is much smaller than the dimension of \sk, which is $4^K$.
 
 The particular torus defined by Eq.~\ref{toro} depends on the Hamiltonian. We may ask how big a space is swept out by
 varying over all Hamiltonians of the form Eq.~\ref{eq:$k$-local}. Specifically does varying over Hamiltonians lead to an almost space-filling  set on \sk?  The answer is no; the number of parameters specifying $H$  (the $J$'s) is polynomial in $K$ and given by Eq.~\ref{miss3}. Thus for a given $k$ the dimension of the set covered by \kl \ evolution is only slightly bigger than a $2^K$-dimensional subset. 
 
 On the other hand we may ask: For each Hamiltonian is the motion on the $2^K$-torus ergodic? Generically the answer is yes. Ergodicity is equivalent to the incommensurability of the energy eigenvalues, a condition  which will be satisfied for almost all members of the ensemble of $J$'s.
 
 To summarize, while the  $\CA$-system is formally defined on a $4^K$-dimensional configuration space, the effective dimension of the system is actually much smaller $\sim 2^K.$

 \bn
 
In Sec.~\ref{sec: randomness} we explained that by starting with a random time-dependent quantum Hamiltonian, a stochastic system can be defined. That stochastic system can be thought of as a classical stochastic version of the auxiliary system $\CA$. Reference \cite{Lashkari:2011yi} refers to such systems as Brownian circuits.  In that case, since the Hamiltonian is now time-dependent, the motion on \sk \ is a random walk not restricted to a torus---it fills up all $4^K$ dimensions and is ergodic on  \sk.

\sc
\section{Geometry of Complexity}\label{sec:geo-comp}

\subsection{The Distance Between Quantum States}\label{sec: distance}
  
Consider the question: how far apart are two quantum states $|A\ra$ and $|B\ra$?   The usual measure of the distance  between them is defined by
\be 
d_{AB}= \arccos|\la B|A\ra|.
\label{distance}
\ee 
The distance $d_{AB}$  is bounded between $0$ (when the two states are the same) and $\pi/2$ (when the two states are orthogonal).
The metric defined by Eq.~\ref{distance} is called the Fubini-Study metric.   It has the property that if $d_{AB}$ is very  small then the expectation values of all observables in the states $|A\ra$ and $|B\ra$  are very close. But this definition misses something important. Suppose we have a very large number of qubits in a complicated pure state that looks thermal, although it is actually pure. Now add one more qubit, either in state $|0\rangle$ or state $|1\rangle.$ Let's call the two states that we get this way $|A\ra$ and $|B\ra. $ They are orthogonal so they are as far apart as possible according to Eq.~\ref{distance}. But in some sense they are not very different; they only differ by the orientation of a single qubit.

On the other hand we can consider two states $|A'\ra$ and $|B'\ra$  in which all of the qubits are mixed up (scrambled) by  two very different scrambling operations. These two states would also be orthogonal, and therefore no further apart than $|A\ra$ and $|B\ra.$ But clearly there is some sense in which $|A'\ra$ and $|B'\ra$ are much further apart than $|A\ra$ and $|B\ra$. The inner product distance of Eq.~\ref{distance}  fails to capture this difference.

The difference between the two senses of distance has operational consequences. Consider the first case with $|A \ra$ and $|B \ra$: it is not hard to create a coherent superposition of states, $\alpha |A\ra + \beta |B\ra;$ nor is it hard to do an interference experiment that is sensitive to the relative phase of $\alpha$ and $\beta;$ and nor is it hard to cause a transition between $|A\ra$ and $|B\ra.$ But doing any of these three things with $|A'\ra$ and $|B'\ra$ would be extremely difficult.

Distances according to the Fubini-Study metric of Eq.~\ref{distance} are conserved under time evolution: the inner product between $U |A \ra$ and $U |B \ra$ is the same as between $|A\ra$ and $|B\ra$. But that does not mean that if they start easy to interfere, they will remain so: large differences between initially similar states can be created merely by the passage of time. Let's take the states $|A\ra$ and $|B\ra$ which are in some sense similar, although orthogonal. Now let's evolve them both by some generic Hamiltonian that allows all the qubits to interact. After a long time the evolved states are 
\bea  
|A'\ra \eq e^{-iHt}|A\ra \cr
|B'\ra \eq e^{-iHt}|B\ra .
\eea
If the system is chaotic then the states $|A'\ra$ and $|B'\ra$ will be very different from one another, and also very difficult to interfere. Some kind of distance between the states will have grown very large. Moreover that distance will continue to grow long after the extra qubit has thermalized with the others. In fact it will grow until it becomes exponentially difficult to interfere the states.

Of course you could argue that the states $|A'\ra$ and $|B'\ra$ are easy to interfere. Just initially interfere $|A\ra$ and 
$|B\ra$ to make $ \alpha |A\ra + \beta |B\ra$ and then evolve the superposition forward for time $t.$ That is true, but the point is that  this way of preparing $ \alpha |A'\ra + \beta |B'\ra$ takes a very long time.  With some locality assumptions we can show that there is no faster way to do it \cite{Aaronson}.

The question then is: Is there a different measure of the distance between states that captures the similarity of $|A\ra$ and $|B\ra$, and at the same time the large difference between $|A'\ra$ and $|B'\ra$? To our knowledge this fundamental issue has not been discussed before. 
Here we propose that the answer is a metric based on a concept of  \it relative complexity. \rm

Consider all the unitary operators that can connect the two states,
\be 
|B\ra = U |A\ra 
\label{B=UA}.
\ee
The relative complexity of $|A\ra$ and $|B\ra$ may be defined as the complexity of the least complex unitary operator satisfying Eq.~\ref{B=UA}. This of course tells us nothing unless we have a criterion for the complexity of a unitary operator. We shall be very brief here and just remind the reader of the concept of circuit complexity. We consider all $K$ qubit circuits composed of \kl \ gates that allow us to prepare $U.$ For simplicity we take the gates to act in series,
\be 
U = g_N g_{N-1}.....g_{1} \, . 
\label{gates}
\ee
The circuit complexity of $U$ is denoted $\CC(U).$ It is the minimum number of \kl  \ gates that it takes to construct $U$ in this way. It depends on the choice of allowable gates; for example it depends on $k,$ but the dependence is rather weak and we assume that it can be accounted for.

We will demand that whenever $g$ is an allowed gate so too is $g^\dagger$; it follows that the complexities of $U$ and $\Ud$ are the same. As a consequence the relative complexity is a symmetric function of $|A\ra$ and $|B\ra.$

Relative complexity defines a notion of distance between states---a complexity metric---which is exactly what we want in order to know how hard it is to make transitions between states, to interfere them, and to measure the relative phases between them in a superposition\footnote{It can be proved that the relative complexity, i.e., the circuit complexity of making a transition between two states; and the circuit complexity of distinguishing the phase of a superposition of the same two states is approximately equal. It can also be proved that the complexity of creating a superposition of the two states is at least as large as the relative complexity \cite{Aaronson}.}.

Relative complexity can also be defined for a pair of unitary operators. Let $U$ and $V$ be such a pair. The relative complexity of $U$ and $V$ is just the complexity of $\Ud V,$ or equivalently $V^{\dag}U.$

Inspired by ideas of Nielsen \cite{Nielsen,Dowling} we will build a new auxiliary theory, $\CA$, based on a complexity metric.

\subsection{Complexity Geometry}\label{sec: right-invariant}

In this subsection we examine and adapt the ideas of Nielsen et~al.~\cite{Nielsen,Dowling} about `complexity geometry'. The idea of the complexity geometry is to make a new metric on SU$(2^K)$, different from the standard metric, in which the distance between two elements of SU$(2^K)$ reflects their relative complexity. We will have a great deal more to say about complexity geometry in a forthcoming paper \cite{UsInFuture}, in which we derive some of the results quoted below, and illustrate them with simple low-dimensional examples. 

There is no  single unique definition of complexity, even in the context of quantum circuits. The definition depends on the allowed set of gates. For example one possibility is to allow all one- and two-qubit gates. Another is to allow up to three-qubit gates, or to choose a discrete collection of gates as long as it is universal. Each gate set gives a different quantitative measure of the complexity of a unitary operator. Since any universal gate set can be simulated by any other universal gate set, the ambiguity is  multiplicative and order unity\footnote{For the purpose of organizing qualitative complexity classes such as $P, \ NP,$  or $BQP,$ which care only about the distinction between polynomial and exponential,  these ambiguities are unimportant.}.

We get a different perspective by focusing on what is not allowed. In this way of thinking we assign a very large, or even infinite, complexity to all unallowed gates. For example we may allow arbitrary gates but penalize all those with weight greater than $2,$ i.e., those involving more than two qubits, by assigning them a large complexity.    This strategy of allowing all gates but introducing a penalty for large gates underlies Nielsen's geometric approach. 

The bottom line is that there is no unique definition of circuit complexity but rather there is a family of complexity measures, which under certain conditions may be multiplicatively related.

\bn

\bn

We need a concept of complexity that is appropriate for continuous Hamiltonian systems, and which matches expectations summarized by the toy model of \cite{Brown:2016wib}.    Nielsen's idea of a geometry of computation---from now on called complexity geometry---is a good starting point.
By a complexity  geometry we will  mean a non-standard metric on \sk \ such that the minimum geodesic distance between points $U$ and $V$ is proportional to the relative complexity (or complexity distance) between them.  Here are some of the features that such a geometry should have:

\bi 
\item It should be a geometry on \sk.
The evolution of $U(t)$ defines a path on \sk. For a discrete quantum circuit the path  consists of a  sequence of discrete segments. The segments  represent individual gates in the case of a series circuit, or  $K/k$ gates for a parallel Hayden-Preskill circuit\footnote{A Hayden-Preskill circuit is one that in each time step applies $K/k$ gates in parallel, so that each qubit is touched exactly once. See Appendix~\ref{sec:terminology}.}.

\begin{figure}[H]
\begin{center}
\includegraphics[scale=.2]{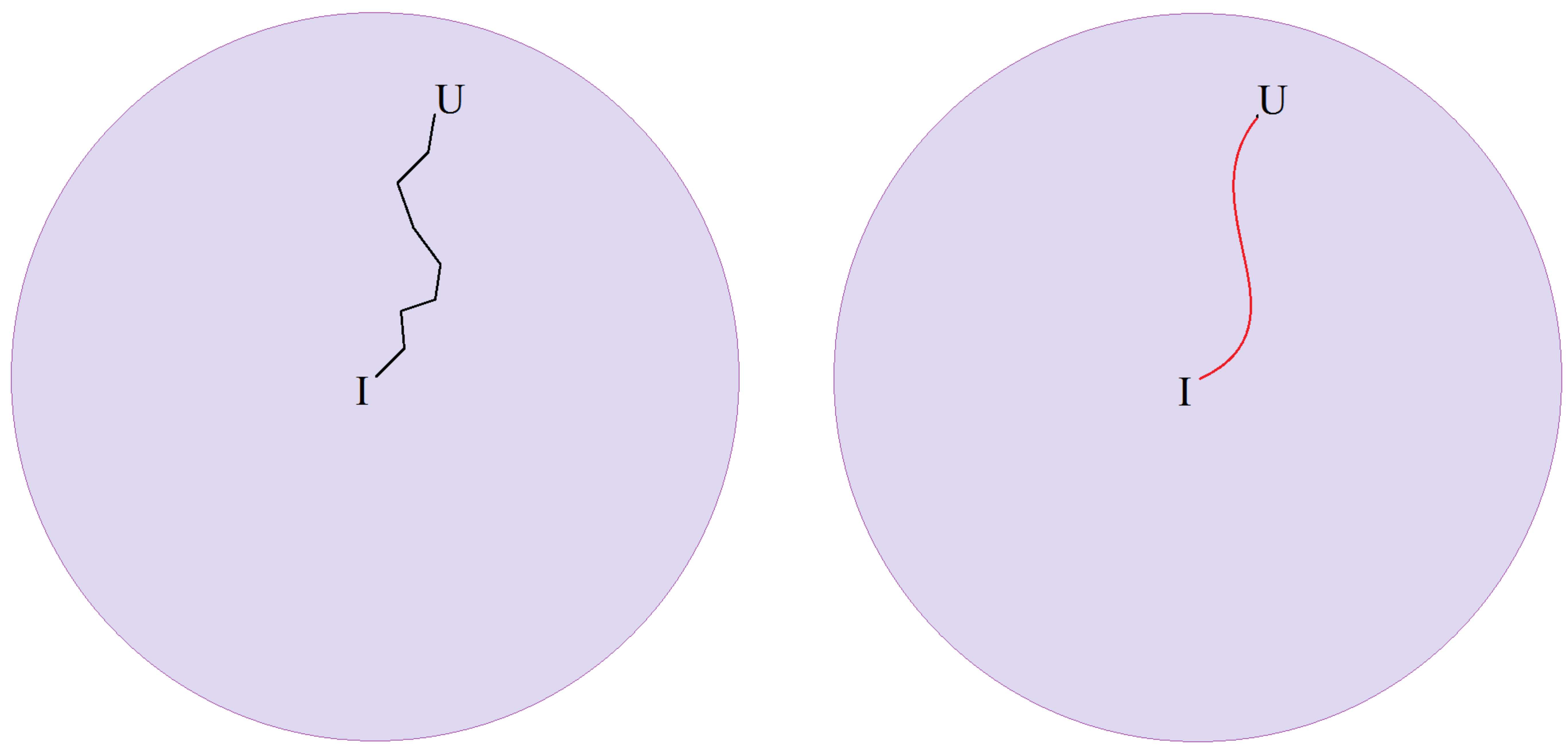}
\caption{The shaded regions in these figures depict the space \sk. The broken trajectory represents the evolution of a discrete quantum circuit and the smooth curve represents Hamiltonian evolution. In both cases computational complexity is identified with the shortest path between the identity and the unitary operator $U.$}
\label{f2}
\end{center}
\end{figure}

For continuous Hamiltonian systems  discrete paths are replaced by continuous paths generated by possibly time-dependent Hamiltonians.  Figure~\ref{f2} shows schematic representations of  discrete and continuous paths through \sk.
\bn

\item
Hayden-Preskill circuits exhibit an effect called the switchback effect \cite{Stanford:2014jda,Susskind:2014jwa}. The switchback effect is closely related to scrambling \cite{Sekino:2008he}.
This same effect appears in the complexity-action duality of  \cite{Brown:2015bva,Brown:2015lvg} and we regard it as an important requirement that the geometry of computation should reproduce it. As we will see Nielsen's original complexity geometry fails in this respect and requires significant modification.

\item The distance function should satisfy the triangle inequality. Given two unitary operators $U$ and $V$ the complexity of the product $UV$ should be less than or equal to the sum of the complexities of the two operators. This follows from the definition of complexity. The triangle inequality is not enough to prove that complexity  geometry is  Riemannian, but we will assume that it is.

\item The geometry should be right-invariant. Consider the construction of $U$ by a sequence of gates in time order starting with the unitary operator $W$:
\be 
U = g_N g_{N-1}.....g_{1}W \, .
\ee
The relative complexity of $U$ and $W$ is the minimum number of gates that satisfies this equation. Now multiply on the right by an arbitrary unitary $V$, 
\be 
UV = g_N g_{N-1}.....g_{1}WV.
\ee
It follows that the relative complexity of $UV$ and $WV$ is the same as the relative complexity of $U$ and $W$. This is not true of $VU$ and $VW.$ 
To see this we write
\be 
VU = (Vg_N g_{N-1}.....g_{1}V^{\dag}) VW \, .
\ee
The operator $V g_N g_{N-1}.....g_{1} V^{\dag}$ will generally not be a product of $N$ gates.
Thus the complexity distance is right-invariant but not left-invariant. Right-invariance is enough to ensure that the  geometry is homogeneous.

\item
All right-invariant metrics are parameterized by a symmetric ``moment of inertia" tensor\footnote{The parallel with the equations for an asymmetric rigid body is intentional. The case $K=1$ is mathematically the same as an ordinary rigid body in three dimensions, since SO(3) = SU(2)$/Z_2$. The matrix $\CI$ would be the moment of inertia tensor and $d\Omega / dt$ would be the angular velocity vector. The symmetric rigid body corresponds to Eq.~\ref{symmetric}. We will have more to say about this in \cite{UsInFuture}.}
 ${{\cal{I}}_{IJ}},$ in terms of which the metric has the form \cite{Milnor}
\be 
dl^2 =  d\Omega_I \ {{\cal{I}}_{IJ}}   \  d\Omega_J ,
\label{right-inv}
\ee
where 
\be 
d\Omega_I = i \ \rm Tr\it \ dU^{\dag}\sigma_I U.
\label{Omega}
\ee

 The metric should penalize motions along directions $\sigma_I$ that  are themselves highly complex. This is the analog of what in circuit theory corresponds to the requirement that gates be simple. Thus we require the metric distance along non-\kl \ directions  to be increased relative to \kl \ directions. This is accomplished by an appropriate choice of  ${{\cal{I}}_{IJ}}.$ The ambiguity in choosing $\CI$ is analogous to the ambiguity in the choice of a gate set for circuit complexity. 

The matrix ${{\cal{I}}_{IJ}}$ should be chosen block diagonal with one block corresponding to the unpenalized \kl \ directions, and the other block corresponding to directions $\sigma_I$ containing more than $k$ single qubit operators. Being unpenalized, the \kl \ block is naturally  taken to be the unit matrix with eigenvalues all equal to $1.$
The non-\kl \ block (the penalized block) should be positive definite, with eigenvalues greater than $1.$ The eigenvalues should increase as the weights of the $\sigma_I$ increase.

\item
It was shown in  \cite{Brown:2016wib} that  consistent descriptions of scrambling and the switchback effect require that  generic   sectional curvatures  be negative, and of order%
\footnote{At first sight this seem inconsistent with \cite{Brown:2016wib} where we claimed the curvature should be $1/K^2.$ The reason for the discrepancy is that in \cite{Brown:2016wib} we assumed complexity is geodesic length rather than action as in Sec.~\ref{C=A}. The factor of $\sqrt{K}$ relating length and action in Eq.~\ref{actandlength} accounts for the difference. This is explained in Appendix~\ref{sec: A vs D}.}
$1/K.$ 
 If no penalty is imposed, in other words if $\CI_{IJ} = \delta_{IJ},$ the metric is  bi-invariant and all sections are positively curved. The introduction of penalty factors tends to make the sectional curvatures negative, but it is not obvious that the natural order of magnitude is $1/K.$  In Sec.~\ref{Sec: sectional curvature} we will show that is indeed the case.
 
\ei

The original version of complexity geometry in \cite{Nielsen}\cite{Dowling} fails badly in this last respect. In our present notation the proposal is equivalent to choosing the non-\kl \ block to be diagonal with all eigenvalues being equal to the enormously large value $4^K.$ As we will see this has the effect of making typical sectional curvatures negative (that's good) but  huge $\sim 4^K$ (not good). This is a far cry from the sectional curvatures $\sim 1/K$ required by \cite{Brown:2016wib} in order to reproduce the switchback effect. The penalty assumed in \cite{Dowling} is much too draconian and must be made more moderate.

\subsection{A More Moderate Penalty}

The reason why \cite{Dowling} chooses the penalty factor for non-\kl \ operators to be of order $4^K$ is that the most complex unitary operators have complexity of order $4^K.$ In order to insure that this is reflected in the properties of the complexity metric, the authors simply penalize all non-\kl \ operators by the common factor $4^K.$ It is certainly true that the  highest weight operators should be penalized by such a factor but the switchback effect requires a much more gradual growth of the penalty as the weight of $\sigma_I$ increases. This will be seen in Sec.~\ref{Sec: sectional curvature}.

Let $w_I$ be the weight of the generalized Pauli operator $\sigma_I.$ We'll assume that the moment of inertia tensor is diagonal,
\be 
{{\cal{I}}_{IJ}} = \delta_{IJ} {{\cal{I}}(w_I)}  \ \ \ \ \ \ \ \rm (no \ sum). \it
\ee

For $w_I 
\leq k$ the coefficients  ${{\cal{I}}(w_I)} = 1 $. Our basic assumption is that the penalty factors ${{\cal{I}}(w)}$ for $w<K$ are independent of $K$. In other words the price that we pay for moving along the direction $I$ is independent of the total number of qubits and depends only on the weight of $\sigma_I.$

For $w_I 
> k$ we assume the eigenvalues smoothly increase from order $1$ to order $4^{K}.$ A simple behavior would be 

\bea 
 {\cal{I}}(w) \eq  1  \ \ \ \ \ \ \ \ \ \ \ \ \ (w\leq k)  \cr 
{{\cal{I}}(w)} \eq c \ 4^{w - k}  \ \ \ \ \ \ (w > k) ,
\label{moderation}
\eea
with $c$ some constant of order one.

We will now show that for\footnote{As an illustration we will consider the case in which there are only weight $2$ operators in the Hamiltonians defining the section.} $k=2$ the typical sectional curvatures are indeed negative and of order $1/K$ as required by the switchback effect.\\
 

In a companion paper devoted to quantitate aspects of complexity geometry \cite{UsInFuture}, we will calculate geometric properties of the complexity metric for various $k$-local systems. In the next section we will show the answer for one particular such system, and show how the sectional curvature is typically negative and of order $1/K$.

\subsection{Sectional Curvature}\label{Sec: sectional curvature}

Our intuition for how complexity geometry should work is based on the two-dimensional toy model of  \cite{Susskind:2014jwa} and   \cite{Brown:2016wib}.
We will now argue that the toy geometry can be thought of as being embedded as a two-dimensional section of the full complexity geometry. 

Certain basic facts about the evolution of complexity, including the switchback effect,  can be summarized by two properties of such sections (that are briefly reviewed in Appendix~\ref{sec: A vs D}):  the typical sectional curvatures are negative; and the magnitude of the sectional  curvature should be of order $1/K.$ 

We will now compute the two-dimensional sectional curvatures of the complexity geometry defined by Eqs.~\ref{right-inv} \& \ref{Omega}, and see that they indeed are negative and of order $1/K.$

Our `section' is the two-dimensional surface consisting of all geodesics through a given point (we will label this point the `origin') that are generated by
linear combinations of two $k$-local Hamiltonians. For definiteness we will choose $k=2$ for the remainder of this section. By definition the sectional curvature is the curvature at the origin.

Without loss of generality we may choose one Hamiltonian to be $H$ and the other to be $H + \Delta d \theta$, where $\Delta$ is a $2$-local operator orthogonal to $H$ and $d\theta$ is an infinitesimal angle.

The surface defined in this way will generally not have zero extrinsic curvature, and so geodesics connecting two points on the section will in general take shortcuts off the surface. For the sectional curvature  defined as the curvature at the origin, $t=0$,  this issue does not arise.

\begin{figure}[H]
\begin{center}
\includegraphics[scale=1.2]{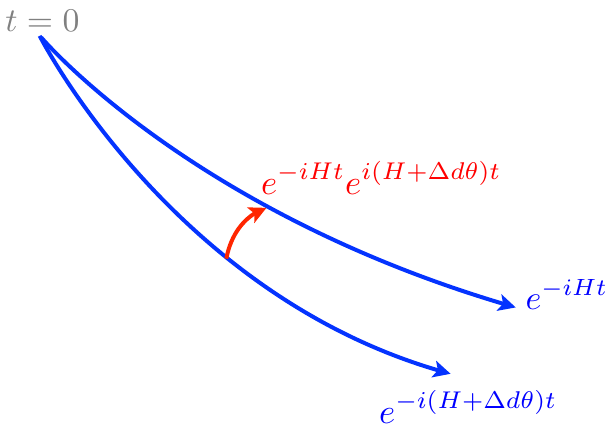}
\caption{Two neighboring geodesics (in blue) leave the origin at $t=0$; this corresponds to evolution under two nearby $k$-local Hamiltonians. The two geodesics are connected by a Jacobi vector (in red); this corresponds to the (non-$k$-local) connecting operator $e^{\Lambda} = e^{-iHt} e^{i(H + \Delta d \theta) t }$. As $t$ increases the connecting Jacobi vector grows, and the acceleration of this growth rate gives the geodesic deviation. In the bi-invariant metric the geodesics always converge, but on the complexity metric the geodesics can diverge for $I_{3}>4/3$.}
\label{f3}
\end{center}
\end{figure}
\bn

\noindent The Loschmidt  echo operator $e^{\Lambda}$ is defined by
\be 
e^{\Lambda} = e^{-iHt}e^{i(H+\Delta d\theta) t} .
\label{Loschmidt}
\ee
By the Baker-Cambell-Hausdorff formula,
\begin{eqnarray}
\Lambda & =&   -  \sum_{m = 0}   \frac{(i t)^{m+1}}{(m+1)!} [\underbrace{H,[H,[H,[H}_{m},\Delta]]]]  \, d\theta \cr   
&=&  -  \left(  i t \Delta   + 
\frac{(i t)^2 }{2}  [ H,\Delta]  + \frac{ (it)^3}{3!}  [H,[{H,\Delta}]] 
 + \frac{(it)^4}{4!}  [H,[H,[H,\Delta]]] + .. \right) d\theta. 
\label{eq:expressionforlambda}
\end{eqnarray}
We find that, to order $d \theta$ and $t^3,$  this comes to
\be 
\Lambda=\left( i\Delta t + \frac{t^2}{2}[H,\Delta] -\frac{it^3}{6}[H,[H,\Delta]] + \ldots \right) d\theta.
\ee
The metric distance along the infinitesimal interval defined by $\Delta$ is
\begin{equation} 
dl^2 =  \Tr  [\Lambda^{\dagger}  \cdot  \Lambda] d\theta^2, \label{eq:dlsquared}
\end{equation}
where the dot-product is taken with the moment of inertia tensor $I_{IJ}$ as the metric\footnote{Let's be more explicit. Since $\Lambda$ is anti-Hermitian it can be written as a weighted sum over the generalized Pauli matrices $\Lambda = \sum_I i \Lambda_I \sigma_I$. With the standard bi-invariant metric the trace would be $\Tr[\sigma_I \sigma_J] = \delta_{IJ}$, but with the complexity metric of Eq.~\ref{right-inv} this trace gets weighted by a factor of the moment of inertia, $\Tr[\sigma_I \cdot \sigma_J] = \mathcal{I}_{IJ} \delta_{IJ}$. Equation~\ref{eq:dlsquared} then gives $dl^2 = \sum_I \Lambda_I^2 \mathcal{I}_{II}$.}. 
The metric distance along the radial direction (the $t$ direction) is
\be 
dl^2 = \Tr [H \cdot H] dt^2.
\ee
In total, to order $t^4$ and order $d \theta^2$ the metric is 
\be 
dl^2 = \Tr [H \cdot H] dt^2+
 \left( \Tr[\Delta \cdot \Delta] t^2 -\frac{t^4}{3} \Tr[ \Delta \cdot [H,[H,\Delta]] ] 
-\frac{t^4}{4}  \Tr [ [H,\Delta] \cdot [H,\Delta]] \right) d\theta^2.
\ee

In order to evaluate the weighted traces, we need to know the weights of the operators involved. The first two terms are easy---since both $H$ and $\Delta$ are by assumption 2-local, and since $2$-local terms by assumption are unpunished, we have $\mathcal{I} = 1$ and so $\Tr[H \cdot H] = \Tr[H^2]$ and  $\Tr[\Delta \cdot \Delta] = \Tr[\Delta^2]$. For the third term, the commutator $[H,[H,\Delta]]$ has both $2$-local and $4$-local pieces, but only the $2$-local pieces survive when the trace is taken against $\Delta$, and so the third term also has $\mathcal{I} = 1$.  The fourth term is harder: $[H,\Delta]$ has both $1$-local and $3$-local pieces. However, as we will argue, in the limit of large $K$ the expression is dominated by the three-local terms, so it is a good approximation to treat it as weighted by a factor we will denote $\mathcal{I}_3$. In total we have
\be 
dl^2 = \Tr [H^2] dt^2+
 \left( \Tr[\Delta^2] t^2 -\frac{t^4}{3} \Tr[ \Delta [H,[H,\Delta]] ] 
-\frac{t^4}{4} \mathcal{I}_3 \Tr \Bigl[ [H,\Delta]  [H,\Delta] \Bigl] \right) d\theta^2.
\label{d(t)}
\ee
Notice that the last term is positive because $[H,\Delta]$ is anti-Hermitian. 

The sectional curvature at the origin ($t=0$) is proportional to the coefficient of the $t^4$ term with a minus sign,
\begin{eqnarray} 
\CR\Bigl|_{t=0, \, K \gg k=2}  & = & \frac{\Tr \left(+\frac{1}{3} \Delta [H,[H,\Delta]]
-\frac{1}{4}{\CI}_{3}[H,\Delta][\Delta,H]\right)}{\Tr \Delta^2 \Tr H^2} \\
& = & \left(\frac{1}{3}-\frac{{\CI}_{3}}{4}\right)    \frac{2 \{ \Tr  [H,\Delta][\Delta, H] \}}{\Tr \Delta^2 \Tr H^2} \, , \end{eqnarray}
where in the last step we have used the identity $\Tr \bigl( [H,\Delta][\Delta,H] \bigl) = \Tr  \bigl( \Delta[H,[H,\Delta]] \bigl)$. The factor in the curly brackets is positive. \\

\noindent We have already assumed that $H$ and $\Delta$ have the exactly 2-local form
\bea  
H\eq \sum J_{ij}^{\alpha \beta} \sigma_i^{\alpha}\sigma_j^{\beta} \cr
\Delta \eq \sum D_{ij}^{\alpha \beta} \sigma_i^{\alpha}\sigma_j^{\beta};
\label{2L}
\eea
let us now assume that all the coefficients are independent random variables with zero mean (for example they could be Gaussian, though that will not be essential). After
averaging over the random variables we have (for large $K$)
\be
\frac{\{ \Tr  [H,\Delta][\Delta, H] \} }{{\Tr \Delta^2 \Tr H^2} }
\sim  \frac{1}{K}.
\ee
Were it not for $k$-locality this quantity would be of order one, but because of $k$-locality all except a fraction $1/K$ of terms in $H$ commute with $\Delta$, so the answer is $\sim 1/K$. Putting it together we will find \cite{UsInFuture}
 \be 
\CR\Bigl|_{t=0, \, K \gg k=2}  =  \frac{2^8}{K}\left(\frac{1}{3}-\frac{{\CI}_{3}}{4}\right) + O \left(\frac{1}{K^2}\right).
\label{important}
\ee

Let us now examine the implications of this remarkable formula. First (as noted in \cite{Dowling}) the sectional curvatures will generically be negative if $\CI_{3}$ is large enough, ${\CI}_{3}> 4/3.$ Second, Equation~\ref{important} shows that if ${\CI}_{3}$ is of order $4^K$ (as assumed in \cite{Dowling}) the curvature will  also be of order $4^K$, which as we explained is too large. In particular it is incompatible with the switchback effect and with the scrambling time being $\log K.$ (Instead, geodesics would deviate so violently that scrambling would be almost immediate.) 
However suppose that Eq.~\ref{moderation} governs the ${\CI}_w.$ In that case ${\CI}_{3}$ is about $4$ and the curvature is of order $1/K.$
 
(We have  confirmed \cite{UsInFuture} that for all \kl \ generalizations of Eq.~\ref{2L} the sectional curvature is generically $\sim k^2/K.$ Again, this is because the probability that two randomly chosen $k$-local terms share a qubit and therefore fail to commute is roughly $k^2/K$.)

The significance of the negative curvature is that geodesics exponentially diverge with typical (local) Lyapunov behavior. This suggests that the motion in complexity space is chaotic. The Lyapunov exponent  corresponds to the exponent that controls the `growth of operators' obtained from out-of-time-order correlations \cite{Maldacena:2015waa}. However, to follow the exponential growth of complexity all the way out to the scrambling time  we would need to compute higher orders in $t$ which we have not yet done.

It should be noted that the sectional curvature at the origin only depends on the first penalty factor $\CI_{3}$. In calculating the geodesic deviation to higher orders in $t$, the higher-weight penalty factors will appear.

\sc
\section{Particle on the Complexity Geometry}\label{sec: particle}
Earlier we considered the motion of a particle on the bi-invariant geometry of \sk. What we really want to study is the motion on the right-invariant complexity metric Eq.~\ref{right-inv}.

What is the relation between the geodesics of the bi-invariant metric and those of the complexity metric  Eq.~\ref{right-inv}? The answer is simple and easy to prove. Suppose the initial velocity components lie in the \kl \ subspace.
In that case the geodesic will always lie in the \kl \ subspace. That follows from the right-invariance of the metric. Furthermore such a geodesic will be exactly the same for either metric---bi-invariant or right-invariant. This includes the length along the geodesic.

In fact we only care about these \kl \ geodesics, but we can generalize the above statement to a wider class. Suppose the initial velocity is any eigenvector of ${\cal{I}}_{IJ},$
\be 
\sum_I \CI_{JI}J_I = \lambda J_J.
\label{eigenvector}
\ee
Again this will continue to be the case along the entire geodesic. Furthermore such geodesics 
 will define  the same curves for both metrics, but the length along them will differ by a factor equal to the corresponding eigenvalue $\sqrt{\lambda}$.

Since we are only interested in the \kl \ geodesics we can calculate them and their lengths from an action principle using either metric. In fact most of the discussion in Sec.~\ref{sec: classical system}  remains the same if we replace the bi-invariant metric with the complexity metric. The only exception is that the complexity metric is not left-invariant and as a result the left charges are not conserved.

\subsection{Geodesic Deviation}
Let us consider a pair of neighboring geodesics, both generated by \kl \ Hamiltonians $H$ and $H+\Delta d \theta$. The geodesics intersect at the origin $t=0$ as in Fig.~\ref{f3}.

Geodesic deviation is defined in terms of  the rate of change of the length of the Jacobi vectors along the geodesics. Because length in the complexity metric and in the standard metric  are not the same, geodesic deviation will be different in the two metrics.
In particular the sign of the geodesic deviation is controlled by the sectional curvature of the section containing the two geodesics. The sectional curvatures in the standard metric are all positive, corresponding to  geodesic convergence (negative geodesic deviation). By contrast  in the complexity geometry the sectional curvatures are typically negative for large enough penalty factors, and geodesics diverge (positive geodesic deviation).
This property of negative sectional curvature is central to the duality between classical and quantum chaos that we are proposing.

Ergodic behavior (see Sec.~\ref{sec:ergodicity})
is necessary for classical chaos but not sufficient. The additional ingredient is the sort of instability characteristic of negative curvature and geodesic deviation.  Without being precise about the definition of chaos, positive deviation leads to the kind of sensitivity to initial conditions that characterizes  chaos.
However, because of the conservation laws the chaos of the $\CA$-system can only take place within a $2^K$-dimensional 
sub-manifold%

The fact that  merely ergodic motion can be made to appear chaotic by changing the metric from the standard  metric to the  complexity metric is a mathematical representation of the differences discussed in Sec.~\ref{sec: distance}. It is something that needs more study.

The negative curvature controls the Lyapunov exponents of the classical auxiliary model and the largest Lyapunov exponent may be identified with the quantum Lyapunov exponent discussed in \cite{Brown:2016wib}.

\subsection{Complexity Equals Action}\label{C=A}
The obvious guess would be that the complexity of the unitary operator $U(t)$
 is the minimal geodesic distance separating it from the identity operator $\mathds{1}$,  
 
\be
\CC = \int_{\mathds{1}}^U \sqrt{G_{MN}dX^M dX^N},
\label{geo-length}
\ee
where the integral is taken along the shortest geodesic connecting ${\mathds{1}}$ and $U.$
However, with the normalization for the metric we have chosen in Eq.~\ref{bimetric}, rather than using geodesic length we instead use  the  action of the  auxiliary system\footnote{This is another example of the connection between action and complexity. Its relation with the action-complexity connection of \cite{Brown:2015bva}\cite{Brown:2015lvg} is at the moment not clear but we certainly find it suggestive.}, as discussed further in Appendix~\ref{sec: A vs D}, 
\be 
\CC = 
\frac{1}{2}\int_{\mathds{1}}^U G_{MN}\dot{X}^M\dot{X}^N d\tau,
\label{quad-action}
\ee
with the constraint that the conserved energy $E_a$ of the auxiliary system is equal to the actual dimensionless Rindler energy $E$ of the quantum system $\CQ,$  
\be
E_a = E .
\ee
Since it is well known (see e.g.~\cite{Susskind:2005js}) that the total energy of a black hole  in Rindler units is proportional to its entropy $K$  we may identify 
\be
E_a = E \sim K. \label{Ea=K/2}
\ee

\bn
Thus we postulate that:

\bn
\it The  complexity of a unitary operator $U$  is the minimum action of any trajectory connecting $U$ and the identity, subject to the condition that the energy $E_a$ of the particle is fixed and equal to $K.$ \rm

\bn
(The minimum action here refers to the action evaluated in the complexity metric, not the bi-invariant metric.)

\bn

The relation between geodesic length Eq.~\ref{geo-length} and the quadratic action of Eq.~\ref{quad-action} is easy to derive,
\be 
\rm Action  = \sqrt{E_a} \  Length \rm,
\ee
or, using $E_a = K,$
\be 
\rm Action  = \sqrt{K} \  Length \rm.
\label{actandlength}
\ee

\bn

Now let us argue that length  and complexity  $\CC$ are related in exactly the same way; in other words that,
\be 
\CC = \sqrt{{K}} \  \rm  Length .
\label{C=sqrtK L}
\ee

In order to normalize length, we  note that according to the standard (bi-invariant) metric the geodesic length between any two orthogonal unitary operators is $\pi/2 \sim 1.$  Along \kl \ directions this is also the distance that it takes for $U$ to become orthogonal to its initial value\footnote{The inner product of two unitary operators $U_1$ and $U_2$ is defined as $\Tr U_1^{\dag}U_2.$ }. In other words  two points $U$ and $U'$ are orthogonal if they are separated along a \kl  \ direction by a distance $\frac{\pi}{2} \sim 1.$

Second, the Aharonov-Anandan bound \cite{Aharonov:1987gg}
  tells us that the time for $U$ to become orthogonal to its previous value (orthogonality time) is $\sim 1/\Delta E$ where $E$ here refers to the $\CQ$ system.  From Eqs.~\ref{E2=J2} \& \ref{J2=K}
 we see that $\Delta E = \sqrt{K}.$
It follows that the orthogonality time is $\sim 1/\sqrt{K}.$ This is discussed in  detail in \cite{Lloyd} and also in \cite{Brown:2015lvg}.

On the other hand the rate at which effective gates act (the rate of complexity growth) is $K$. Therefore the number of gates that act in  an orthogonality time is $\sim \sqrt{K}.$ Putting it all together we see that the number of gates corresponding to a geodesic distance  $\sim 1$ is $\sqrt{K}.$ Thus  the complexity accumulated over a  distance $L$ is 
\begin{equation} 
\Delta \CC = \sqrt{K} \Delta L,
\end{equation}
where $L$ is length. 
It follows that complexity and length differ by precisely the same factor---namely $\sqrt{K}$---as action and length. The factor of $\sqrt{K}$ in Eq.~\ref{C=sqrtK L} is the same factor that appears in Appendix 
\ref{sec: A vs D}.

\subsection{The Growth of Complexity}

From the ordinary nonrelativistic  connection between  kinetic energy and velocity, and from Eq.~\ref{Ea=K/2}, we find that the velocity of the auxiliary particle satisfies,
\be 
L_a = \frac{1}{2}V_a^2 =  K .
\ee
It also follows that the value of the Lagrangian is
\be  
L_a= K .
\ee
Our basic hypothesis---that complexity equals action---implies that the rate of growth of complexity is $L_a$. Thus we find that as expected, complexity grows according to
\be 
\CC = Kt.
\ee

\bn

As we have seen, the classical motion generated by \kl \ Hamiltonians takes place on a sub-manifold of dimension slightly larger than $2^K.$ 
Recall the conjecture of  Sec.~\ref{Sec: QC and CE}  that the quantum computational complexity of a $K$-qubit system $\CQ$ evolves in a similar manner to the entropy of  a classical system with $\sim 2^K$ degrees of freedom. The new conjecture should now be obvious: up to a factor to be determined, the ensemble-averaged complexity of $\CQ$ is the entropy  of $\CA,$  denoted $S_a.$  We will have to refine this idea, but in essence that's the proposal. Given an energy and  entropy, the classical  system $\CA$ has its own thermodynamics which is quite distinct from the thermodynamics of $\CQ.$ We may call it the thermodynamics of complexity.

As we mentioned at the end of Sec.~\ref{sec: Qsystem} we can generalize the quantum system  by allowing stochastic time dependence in the $J$'s. The effect on the classical auxiliary system is to turn it into a problem of diffusion on the complexity geometry.

\sc
\section{Statistical Mechanics of Complexity}\label{sec: stat-mech}

As we mentioned early in this paper, the growth of complexity for a quantum system of $K$ qubits resembles the growth of entropy for a classical system with an exponential number of degrees of freedom. We will now consider the statistical mechanics of $\CA$ and how it is related to the complexity of $\CQ.$

\subsection{Average Complexity Equals Entropy}
The phase space probability distribution for a classical non-relativistic gas often separates into two factors, one depending on the positions of the particles and the other on the momenta,
\be 
P(x,p) = F(x) G(p).
\ee
As a consequence the total entropy is a sum of two terms: the positional entropy associated with the distribution $F(x),$ and the \it kinetic \rm entropy associated with $G(p),$ 
\be 
S= - \int  F(x)\log{F(x)} \ dx -\int  G(p) \log{G(p)} \ dp.
\ee
It is not necessary that the system be in equilibrium for the entropy to separate in this way. It is only necessary that the probability factorizes.

Let us now state the basic two-part conjecture.  The first part is about the computational complexity of $U$ and the positional entropy of the $\CA$-system. The second part is about kinetic entropy and Kolmogorov complexity. In both cases the term \it ensemble average \rm implies an average over initial velocities, or by V/J-correspondence, an average over the couplings $J.$

\subsection{Computational Complexity and Positional Entropy}
Our conjecture states that:

\bn
\it 
At any instant, the ensemble average of the  computational complexity of the quantum system $\CQ,$
 is proportional to the classical positional entropy of the auxiliary system \sa.
\rm

\bn
There are two qualifications to note. The first is that we identify computational complexity with
 \it  positional entropy \rm   instead of total entropy. The reason for this qualification
is that computational complexity has only to do with the distance of a point $U$ from the origin; in other words its position in complexity space, not its velocity. In subsection~\ref{KC_and_KE} we will consider kinetic entropy and its connection with complexity.

The other qualification is  the use of \it proportional to \rm rather than equal to. Computational (or circuit) complexity depends on a number of factors such as the gate set. We assume that different choices lead to a multiplicative ambiguity in the definition of complexity. On the other hand, if our conjecture is correct,  a particular normalization of complexity  will allow us to equate average complexity with positional entropy. 

The conjecture can be stated in another way. We consider the number of unitary operators that can be reached by time-independent \kl \ Hamiltonians, with complexity less than or equal to $\CC.$ Call it $N(\CC)$. Our conjecture amounts to the claim that for $1\ll\CC < 2^K $
\be  
N(\CC) = e^{a \, \CC}, \label{N=exp C}
\ee
with $a$ being a constant independent of $K$, but dependent on the specific scheme (gate set, etc.) for defining complexity. If true it would allow us to define a normalization for complexity for which $a=1.$

An intuitive counting argument for the conjecture will be given shortly.

\bn

\subsection{Kinetics}

We have considered the positional aspects of entropy. Now let us consider the kinetic aspects.
The
 auxiliary energy in Eq.~\ref{E-sub-a} is simply expressed in terms of $J,$ 
\beq
E_a = \frac{1}{2}\sum J^2,
\ee
i.e. the energy is proportional to the sum of the squares of the couplings. 
Recall that the probability distribution $P(J)$ in Eq.~\ref{P(J)} has the form of a Gaussian. Using the velocity-coupling (V/J) correspondence of Sec.~\ref{sec: VC}
this distribution is seen to be a Maxwell-Boltzmann velocity distribution,
\be 
P(V) = \frac{1}{Z}e^{-\frac{1}{2}B_a \sum_I V_I^2}.
\label{MaxBolt}
\ee
Alternatively it defines  a Gibbs ensemble,
\be 
P= \frac{e^{-B_a E_a}}{Z},
\ee
with the constant $B_a$ being the inverse temperature of the auxiliary system.
\be 
T_a = 1/B_a.
\ee

The temperature may be determined in a number of ways, the easiest being to use the fact that every degree of freedom in a Maxwell-Boltzmann distribution has energy $T_a/2.$ The result depends on the locality parameter $k.$ For illustration we consider $k=2.$
The total energy is given by Eq.~\ref{Ea=K/2} as $E_a = K$ and there are of order ${K \choose 2} 3^2$ excited degrees of freedom\footnote{Once again the factor $9=3^2$ is due to the three Pauli operators for each qubit.}. Thus the energy per degree of freedom is $2/9K$ and the temperature is 
\be 
T_a = \frac{4}{9K}.  \label{eq:Ta2over9K}
\ee

More generally, if the Hamiltonian is \kl \ instead of 2-local the temperature will satisfy,
\be 
 T_a \sim 1/K^{k-1}.
 \ee
(To be clear, $T_a$ is the temperature of the classical auxiliary system $\CA$; it is not the temperature of the quantum system $\CQ$.)

There is of course an entropy associated with the probability distribution of the $J$'s. By  V/J-correspondence it may be thought of as the kinetic part of the total entropy.

\subsection{Kolmogorov Complexity and Kinetic Entropy}\label{KC_and_KE}

This raises an interesting question: What, if anything,  does the kinetic term in the entropy have to do with complexity? It would be odd and maybe disappointing if one term in the auxiliary entropy (the positional entropy) was an average complexity and the other (the kinetic entropy) was not. We don't believe  this to be the case.

Given that the velocities are related to the $J$-coefficients we can identify the kinetic entropy of the classical auxiliary system $\CA$ with the entropy of the probability distribution $P(J).$ (Since this is a function of the coupling constants $J$, this is a property not of the quantum \emph{state} but of the quantum \emph{Hamiltonian}; this is a consequence of Sec.~\ref{subsec:equationsofmotion}, where we saw that the velocities in $\CA$ are given by the quantum Hamiltonian in $\CQ$.)

For a moment suppose the $J$'s are each either $0$ or $1.$ The Hamiltonian Eq.~\ref{eq:$k$-local} would then be specified by a bit-string $(0110100.....)$. It would be natural to ascribe a Kolmogorov  complexity $\CC_{\kappa}(s)$ to the string $s$. 
Kolmogorov complexity measures the length of the shortest algorithm that can prepare a string. Applied to the string of $J$'s it would define a Kolmogorov complexity for each specific instance of a Hamiltonian.

The Kolmogorov complexity is a measure of randomness which, unlike classical entropy, does not depend on probabilistic assumptions, or the existence of a statistical ensemble. In some respects it is a more physical quantity than entropy in that it is defined for each instance and  does not make reference to the state of knowledge of the observer \cite{Zurek:1989zz}.
 Its disadvantage is that it is uncomputable and difficult to work with. Fortunately under suitable assumptions the \emph{average} Kolmogorov complexity \emph{is} connected to entropy.

If we are given a statistical ensemble of bit-strings we may define two measures of randomness or genericity for the ensemble. The first is the good old entropy defined  by the usual formula $-\sum P(J) \log{P(J)}$.  The other is the ensemble average of the  Kolmogorov complexity $\sum P(J) \CC_{\kappa}(J)$. What if anything is the relation  between these quantities?
In fact under mild assumptions\footnote{The important assumption is that the probability distribution itself not be too complex. For  simple distributions, such as Gaussian, this complexity is negligible. See Grunwald and Vitanyi \cite{Grunwald}.} the two are the same \cite{Zurek:1989zz}\cite{Grunwald},
 \beq
-\sum P(J) \log{P(J)} = \sum P(J) \CC_{\kappa}(J) = \langle \CC_{\kappa} \rangle.
\ee

The $J$'s are real numbers, not binary digits. This means that to specify them with infinite accuracy will in general take an infinite amount of information, which means infinite Kolmogorov complexity (the same infinity that shows up in the classical entropy of continuous variables, such as velocity).  We will fall back on a discrete approximation to the continuum. For example, suppose $J$ takes on real values on some interval. We can replace the real numbers by a fine 
lattice with spacing $\delta.$ All together there are $\sim 1/\delta$ points on the lattice. A value of $J$ can be specified by an integer from $1$ to $1/\delta.$ It is well known that the typical Kolmogorov complexity of such an integer is of order $\log{(1/\delta)}$ and therefore diverges logarithmically as $\delta \to 0.$

Despite this divergence we still expect the ensemble averaged complexity to be the same as entropy. This is because the same $\log \delta$ divergence appears in the entropy for the reason that the probability for any value of $J$ is order $\delta$ and the sum $\sum P \log P$ will be proportional to $-\log \delta$.
The average Kolmogorov complexity of the $J$'s depends logarithmically on the tolerance in specifying  the Hamiltonian\footnote{To be clear, we are calculating the Kolmogorov complexity of a time-independent Hamiltonian, with a tolerance $\delta$. The time $t$ does not appear. If instead we were calculating the Kolmogorov complexity of a quantum state evolving under a time-independent Hamiltonian---which, to be even clearer, is not the quantity of interest for the purposes of relating $\CQ$ and $\CA$---we would find that this generically scales like  like $\log t$ at intermediate time. Consider the algorithm for specifying the state that first gives the (simple) initial state, then says `evolve for time $t$', and then specifies the time-independent Hamiltonian to be used in the evolution. The first part is a fixed overhead that doesn't scale with $t$. The second part---specifying the time---requires $\log t$ bits. The third part---specifying the Hamiltonian---also requires $\log t$ bits, because to approximate $e^{-iHt}$ for a time $t$ requires an accuracy in $H$ that is an inverse polynomial in $t$ \cite{Loschmidt}. \label{footnote:Kolmogorov}}.

\bn
Before concluding this section we will give a circuit analogy.
The analog of Hamiltonian evolution would be to start with a unitary circuit of small depth, call it $\Gamma,$ and to repeat it over and over,
\be 
U(t) = \Gamma^t \ \ \ \ \ \ (t \ = \ \rm integer\it ) .
\ee
Most of what we described here can  be adapted to that case. Averaging over Hamiltonians would be replaced by averaging over an ensemble of $\Gamma$'s.  

In computer science terms $\Gamma$ is the  program that determines what computation the circuit carries out. Part of the complexity of the entire computation is the Kolmogorov complexity of $\Gamma.$ The ensemble average defines an entropy, which as we've seen, is related to  kinetic entropy.

The kinetic entropy of the $\CA$ system is time-independent and of order the number of $J$'s. This is polynomial in $K.$ On the other hand the positional entropy is time dependent and can grow to exponential size $\sim 2^K$ at equilibrium. During the early period of complexity growth the two can compete but in equilibrium the entropy is dominated by the positional term. To put it another way, the Kolmogorov complexity is essentially a fixed overhead having to do with the complexity of the algorithm, but after the algorithm has run for a long time the computational complexity vastly exceeds the fixed overhead.

The computational complexity measures the total number of gates required to build the minimal circuit that generates the state. Even for a time-independent Hamiltonian, this scales like $t$ since you need to keep paying over and over again to apply the same gates over and over again. The Kolmogorov complexity is (no more than) the number of bits in the most compressed possible description of this circuit. For time-independent Hamiltonians you do not need to keep paying over and over again as you concatenate identical sub-circuits, since you can just specify the total number of such sub-circuits with $\log t$ bits.

Whether or not we add the Kolmogorov complexity to the circuit complexity to define a total complexity is a matter of definition. In ordinary thermodynamics the two kinds of entropy are transmutable into each other, for example by adiabatic compression or expansion, so adding them is essential. In the present context one thing is clear: the Kolmogorov complexity of $\Gamma$ and the circuit complexity \emph{both} contribute to the overall complexity of carrying out a computation.

\subsection{A Counting Argument}\label{sec: counting}
The set of operators reached by evolving with \kl \ Hamiltonians forms a space of dimension not much bigger than $2^K.$ Ideally  we would like to know how much of the volume of that space is occupied by operators of complexity $\CC.$ The conjecture of Eq.~\ref{N=exp C} is that it is exponential but we haven't proved it. Brownian or random circuits which fill all $(4^K-1)$ dimensions of \sk \ are easier to analyze.
  The counting problem in this case is the unrestricted counting of all unitary operators with complexity less than or equal to $\CC.$ We'll do that counting now.

There is an important difference between the time-independent Hamiltonian model and stochastic random circuit models. The difference has to do with the Kolmogorov complexity of the circuit.
In both the time-independent Hamiltonian model, and the repeated-$\Gamma$ model the Kolmogorov complexity is essentially a fixed overhead which does not grow linearly as the circuit evolves. Thus whether we include it or not, we may ignore it over long timescales. 
This is not the case for random circuits, where at each time step  a new random choice of gates has to be made. It is evident that the Kolmogorov complexity increases linearly with time and therefore, if included, it will contribute to the growing total complexity of an evolving circuit. In what follows we include the Kolmogorov complexity in the counting for a stochastic or Brownian circuit.

In the  simplest model  at each instant a single gate acts.
If we only have to choose from a small gate set, the Kolmogorov complexity per gate would be order $1$ and would not be very important. But at each step the choice also involves which set of $k$ qubits the gate acts between. For example in the case $k=2$ there are $K(K-1)/2$ possibilities to choose from. That means that each gate adds a Kolmogorov complexity $\sim \log{K^2}.$ We can easily account for this by  assigning a complexity $\log{K^2}$ to each gate. (We would not do this in the $\Gamma$ model in which the Kolmogorov complexity is essentially a fixed overhead. In that case each  gate 
is assigned complexity $O(1)$.) The full complexity of a unitary operator in the stochastic model is $\log{K^2}$  times the minimum number gates that are required to prepare $U.$

\bn

We can give a rough counting argument  for how complexity grows. The argument is closely related to one                                                                                                                                                                                                                                                                                                                                                               given by Roberts and 
Yoshida \cite{Beni}. Let's consider a path through $\suk$ defined by a series of  $n$ 2-qubit gates
\be 
U(n) = g_n g_{n-1}...g_1.
\ee
The gate-set is assumed universal and includes $m$ gate-types which  can  act on any pair of qubits. Thus each gate involves a choice of $ \frac{ mK(K-1)}{2}$ possibilities. The system of paths defined this way forms a tree \cite{Brown:2016wib}.  The tree is a  discrete analog of complexity geometry.

The number of such paths of  length $n$ is 
\bea
N(n) &\sim & \left(\frac{ mK(K-1)}{2}  \right)^n
\cr \cr
&\sim& e^{n  \log{(mK^2)}}.
\eea
Does each path produce a different unitary operator or are there collisions where two paths produce the same operator? Because of the  very high dimensionality of $\suk$ collisions of this type are rare until $n$ is very large. In fact the fundamental assumption that this work is based on is that collisions do not generically occur until $n$ is exponential in $K.$ 

Under these conditions the set of unitary operators that can be reached in this way include all $U$ with complexity less than $\CC=n\log{K^2},$ and no $U$ with complexity greater than $n\log{K^2}$. The conclusion is that the number of unitary operators with complexity less than or equal to $n\log{K^2}$ is $N(n).$ Thus the number of $U's$ with complexity of order $\CC$ is
\be 
N(\CC) = e^{ \CC}.
\label{overestimate}
\ee

We may think of all the operators with complexity between $\CC$ and $\CC + \delta \CC$ as living in a shell of volume $e^{\CC}$ surrounding the root of the tree. The positional entropy of an ensemble supported in this shell is the logarithm of this volume and is therefore
\be 
S_a\approx \CC .
\label{S-approx=C}
\ee
 
Thus, if the Kolmogorov complexity is included,  in the stochastic model we are justified in identifying average complexity with auxiliary entropy.\\

This counting argument relies on the assumption that at subexponential times collisions are rare. This assumption seems particularly warranted in the context of the stochastic random circuit model we have considered so far in this subsection. To establish Eq.~\ref{N=exp C} we'd like to make the same assumption in the context of unitary operators generated by the exponentiation of $k$-local time-independent Hamiltonians. In this case the assumption seems less secure, since by restricting ourselves to this special subset of unitaries we may have made collisions more likely. Nevertheless,  the subset of unitaries that may be generated by $k$-local time-independent Hamiltonians is still exponentially big, and our conjecture is that this should be good enough to underwrite Eq.~\ref{N=exp C}.

\subsection{A State-Complexity Argument}\label{sec: state-complexity}

We'll give one more argument for   Eq.~\ref{N=exp C}, not based on operator complexity but on state complexity.
Earlier, in Sec.~\ref{sec: distance}, we discussed relative state complexity. In order to define absolute state complexity one needs a concept of a simple state. By a simple state we will mean one with no entanglement, for example the product state $|000..00\ra.$ Once one specifies what states are simple, the absolute state complexity of $|\psi\ra$ just means the smallest relative complexity between $|\psi\ra$ and a simple state. To say it another way, the state complexity of $|\psi\ra$ is the minimum number of gates required to convert it to an unentangled state.

 The geometry of state complexity is similar to that of unitary operator complexity \cite{UsInFuture}.  The most important difference with unitary operator complexity is that the space \sk \ is replaced by the projective space of normalized states $CP(2^K-1).$

In order to count states we have to regularize $CP(2^K-1)$  by dividing it into cells of linear size $\epsilon.$ The number of such cells in $CP(2^K-1)$ is obtained by dividing the volume of $CP(2^K-1)$ by the volume of a ball of radius $\epsilon.$ The answer is that the number of states is given by
\be
N_{\epsilon}=\epsilon^{-2^K} = e^{|\log \epsilon|2^K}.
\label{log-div}
\ee
 This is often simplified to $e^{2^K}.$ We will return to the $\epsilon $ dependence in a moment.

Now consider the number of states that have complexity $\CC.$ Let us assume that it is exponential
\be
N(\CC) = e^{\alpha \CC},
\label{N=eaC}
\ee
where $\alpha$ is a constant to be determined. The maximum state complexity is $\sim 2^K$  and almost all states have that complexity. On the other hand the total number of epsilon-balls in $CP(2^K-1)$ is given by 
\be
N_{\epsilon}= e^{2^K}.
\label{Nepsilonball}
\ee
 Consistency of Eq.~\ref{N=eaC} and Eq.~\ref{Nepsilonball} requires $\alpha =1.$ Thus the number of quantum states with a given complexity grows as the exponential of the complexity. Taking the logarithm implies that average complexity is auxiliary entropy.
 
 Coming back to the $\epsilon$ dependence, the logarithmic divergence in the counting of states is familiar in classical statistical mechanics. Strictly speaking the continuous nature of phase space implies that entropy is infinite. The divergence may be regulated by discretizing space and momentum space, and one finds the divergence being logarithmic as in the exponent of  Eq.~\ref{log-div}.

On the complexity side we have been a bit sloppy  
 in claiming that the maximum complexity is $2^K$.  Complexity, like entropy, also requires a cutoff, and a more correct statement is that the maximum complexity is $| \hspace{-.5mm} \log{\epsilon}| 2^K.$  Thus the divergences in complexity and entropy match.\\

The two arguments we've given---the counting argument and the state-complexity argument---are arguments for the plausibility of our conclusion, but are far from rigorous, and it would be interesting to explore this question further. 

\sc
\section{The Second Law} \label{sec: second-law}
In this section we come back to the original question that we asked in Sec.~\ref{Sec: QC and CE}: is there a Second Law of Complexity? Let us first discuss an obstruction to complete thermalization of $\CA.$

\subsection{Obstruction to Thermalization}
The Maxwell-Boltzmann velocity distribution  in Eq.~\ref{MaxBolt}  is an initial condition connected with a choice of a Gaussian distribution for the coupling constants $J.$ It is not a consequence of dynamical thermalization of the $\CA$ system. In fact the large number of conservation laws associated with right-multiplication invariance
creates an obstruction to thermalization. 
By contrast the tendency toward maximal positional entropy is not obstructed and takes place for each value of the conserved quantities.

There are $4^K$ conserved generators of right multiplication. They are given by Eq.~\ref{Gen=sUU}. Within each leaf of the foliation (by the values of the generators) the auxiliary system with complexity metric is chaotic. This means that the positional entropy will grow with time and eventually reach its maximum, but if the initial velocities are not Maxwell-Boltzmann distributed, the system will never reach thermal equilibrium. This is roughly like a gas of completely free particles on a very large negatively curved Riemann surface. The kinetic energy  of every particle is conserved, but  the positions will spread out and eventually fill the space.

 Granting the correspondence between average complexity and auxiliary entropy, we can  give a rough analogy for the growth and evolution of computational  complexity. Initially a large number $\sim 2^K$ of particles are located near the origin of a large box of volume $e^{2^K}.$ The  velocities are Maxwell-Boltzmann distributed. The gas begins to expand and the positional entropy grows. Eventually  the gas fills the box and  comes to equilibrium. It stays in equilibrium for a very long time but on timescales $e^{2^K}$ recurrences happen. Figure~\ref{f1} is the result of translating this picture into the computational complexity of the system $\CQ$.

\subsection{Second Law of Complexity}

The thermodynamic laws of complexity are just the usual laws of thermodynamics applied to $\CA.$ The second law, applied to positional entropy implies a second law of computational complexity \cite{Brown:2016wib}:

\bn

\it
If the computational complexity is  less than maximum, then with overwhelming likelihood it will increase, both into the future and into the past. \rm

\bn

The classical system $\CA$ tends to positional  equilibrium after a time polynomial in the number of classical degrees of freedom, and  then remains in equilibrium for a classical recurrence time. This implies that the quantum system $\CQ$ comes to complexity equilibrium after a time exponential in the number of qubits, and remains there for an even greater quantum recurrence time, the quantum recurrence time being doubly exponential in $K$. 
Thus we achieve our goal of understanding the growth of complexity for a $K$ qubit $\CQ$ system (Fig.~\ref{f1}) in terms of the behavior of classical entropy for an $\CA$ system of $2^K$ degrees of freedom.

\bn

In principle one can reverse the evolution of a large but finite system by intervening with a process which changes the sign of its Hamiltonian. In classical physics this reverses the trajectory in phase space and if it can be done with sufficient precision it will reverse the increase of entropy, causing an apparent violation of the second law of thermodynamics. The only problem is that  decreasing entropy is unstable when the system is chaotic. The effect of a tiny change in a single degree of freedom will exponentially grow, and quickly reverse the decrease of entropy, turning it back to an increase. 

We can apply this property of classical physics to the $\CA$-system and derive an important property of quantum complexity.
In principle quantum states of a many body system can be prepared which will evolve toward decreasing complexity\footnote{In gravitational physics a white hole is an example of a state of decreasing complexity.}. But the quantum-classical duality between system $\CQ$ and system $\CA$ implies that the decrease is unstable. The application of a small perturbation to a single degree of freedom will exponentially spread through the system, and reverse the decrease of complexity. This phenomenon and its relation to negative curvature was studied in the toy model \cite{Brown:2016wib}. It can also be 
 explicitly seen in black hole dynamics using the classical shock wave calculus of 
\cite{Shenker:2013yza}.

Finally, as pointed out in \cite{Brown:2016wib}, the largest classical Lyapunov exponent of $\CA$ is the quantum Lyapunov exponent \cite{Maldacena:2015waa} of $\CQ.$

\sc
\section{Uncomplexity as a Resource} \label{Sec: combining}

When a classical gas explodes from the corner of a room that does not contain a turbine, the increase in entropy is wasted.


In a similar way, when a black hole evolves, it uselessly generates complexity.  Black holes are not only the fastest computers in nature, they are also the most useless. They implement the highest number of gates per unit mass per unit time, but which gates they implement are chosen by quantum gravity, not by the user. The result is computation that, while extremely fast, is undirected---useful only for those whose purpose is to simulate black holes. 



But the second law of thermodynamics has another side to it beyond the inevitability of the increase in entropy, the side that led to its creation by steam-engineers. An entropy gap, namely the difference between the entropy of a system and the maximum entropy in thermal equilibrium, is a resource \cite{Spekkens}. This resource can be harnessed to perform directed work. 

In this paper we are interested in the question of whether complexity defines a resource that can be harnessed in a useful, directed, manner, in analogy with thermodynamic work. We expect that the analog of directed work is directed quantum computation---we will call this `computational work'. 

In exploring this conjecture, we will be guided by the analogy between the complexity of the quantum system $\CQ$ and the entropy of the classical system $\CA$. This is an incompletely-defined idea, but nevertheless we will give some reasons to believe that a resource interpretation of complexity exists.\\

Without giving a formal definition of thermodynamic `work', for a process to do work it must have the following features: 
\begin{enumerate}
\item Doing work enacts a directed transition from one macroscopic state to another, with a deliberate goal. (For example, raising a weight.) 
\item Doing work expends a resource. Once the available resource is fully expended, no further work is possible until the resource is replenished. 
\item Doing work involves a procedure that depends only on the macrostate of the system involved, and not on the specific microstate.
\end{enumerate}
(This definition of work excludes the kind of work that involves Maxwell's Demons.)\\

\bn

By a quantum computation we will mean a quantum circuit that begins with a pure input quantum state and ends with a pure  output quantum state. The circuit may be composed of gates or a possibly time-dependent Hamiltonian. In other words it is a quantum-in---quantum-out process and its purpose is to reach a target state.  The computation  can be thought of as a trajectory on the space of states or in the configuration space of the auxiliary system $\CA$. No measurement  is allowed during the course of the computation, as measurements are not part of the $\CQ$-$\CA$ correspondence.
Of course to be useful the computation must be followed by a measurement but only at the very end.  The computational work and the necessary resources  refer to the quantum-in---quantum-out computation  and not to the measurement.

In thermodynamics, the free energy
 $$F=E-TS,$$
is a resource that represents the amount of energy that can be directed toward useful work. Applied to the auxiliary system the definition of free energy would be $E_a -T_a S_a$, or equating auxiliary entropy with complexity,  
 \be 
 F_a= E_a - T_a \CC.
 \ee
 For the auxiliary system, as formulated thus far, both the energy $E_a$ 
 and the temperature $T_a$ are fixed parameters that only depend on the number of qubits through Eqs.~\ref{Ea=K/2} \&~\ref{eq:Ta2over9K}. The only variable in the free energy is the complexity. Therefore we propose that the quantity $-\CC$ be treated as a resource. More exactly we propose that the gap between the complexity and the maximum possible complexity---the `uncomplexity'---is a resource that can be utilized for directed computation, 
 \be 
\textrm{Resource} = \Delta \CC = (\CC_\textrm{max} - \CC ). \label{eq:definitionofuncomplexity}
 \ee 
 
 To understand why uncomplexity might be viewed as a resource, let's consider how useful a computer would be if the resource is all used up. Consider a tired old  quantum computer that has been allowed to run for such a long time that the state-complexity has reached its maximum value, exponential in $K$, and therefore $\Delta \CC=0.$
 
For most purposes a state of maximal complexity is indistinguishable from a maximally mixed density matrix. In both cases the expectation values of all but the most complex operators  are given by their Haar-random values. Suppose our computer is initialized in a mixed state with density matrix proportional to the unit operator, 
$$
\rho \sim {\mathds{1}}.
$$ 
Consider any unitary operation $G$ that we may apply. (We use the notation $G$ to suggest that the operation may be composed of gates.)  The action of $G$ on any density matrix changes it to $G \ \rho \  G^{\dag}.$ This may or may not be useful in general, but when applied to the maximally mixed density matrix it does nothing. Whatever operation is applied, the result is the same: the maximally mixed state. Therefore unless the computer is re-initialized no useful computation is possible.

The same is true for a maximally complex state as long as $G$ is not so complex that it can undo the exponential complexity of the initial state.

The state with the maximal resource has $\CC=0$ which means a simple  unentangled product state. It seems reasonable that the most powerful initial state for general all-purpose computing would be the simplest state.

In attempting to think of the uncomplexity $\Delta \CC$ as a resource we will use the correspondence between the quantum complexity of $\CQ $ and the classical entropy of $\CA$ as a guide.
We will now give some examples based on thermodynamic analogies.

\subsection{Combining Systems:  A Paradox}\label{Sec: combining 2}

Many thermodynamic questions concern what happens when  two isolated systems, each in equilibrium, are brought into contact. The first question  is: What does it mean to combine two auxiliary systems, and how is it related to combining the corresponding quantum systems?
Here we will consider a simple case: two thermodynamically identical $\CA$ subsystems at the same temperature $T_a$ and entropy $S_a$ are combined. This   should give rise to  a single system in equilibrium  at the same temperature, with an entropy $2S_a.$

We would like to understand what it means to combine two classical auxiliary systems, each in complexity equilibrium, into a composite auxiliary system. In other words given an auxiliary system $\CA,$ what is the meaning of $\CA \times \CA$?

Here is the paradox: Naively we might think that combining two auxiliary systems involves combining the two corresponding quantum systems in the form $\CQ\otimes \CQ$, where each  factor contains $K$ qubits, and is in complexity equilibrium.
 Let's see what happens if we do so. Each subsystem has complexity of order $\CC=2^K.$ 
Immediately after combining the systems the total entropy is $2\times 2^K$, and the maximum complexity of the combined system is 
\beq
\CC_\textrm{max}= 2^{(2K)}.
\ee
This is the square of the individual complexities, not the sum. Therefore the resulting systems, when combined, will be very far out of complexity equilibrium. That is not what should happen if we combine two identical thermodynamic systems; the entropy should be additive.

Evidently combining two quantum systems does not correspond to combining the auxiliary systems in an additive way. Instead it multiplies the number of degrees of freedom of the auxiliary systems. This seems to be evidence that complexity does not behave like entropy. 

The resolution of this paradox is that the operation 
of  combining  auxiliary systems is entirely different from combining the corresponding quantum systems. The right idea is to take the system of $K$ qubits and \it add just a single additional qubit.\rm \  Adding one qubit doubles the dimension of the Hilbert space, and therefore doubles the number of classical degrees of freedom of the auxiliary system.

\bn

\bn
 
Let's show this in equations. If $|\psi_0 \ra$ and $|\psi_1 \ra$ are both $K$ qubits states with $\la \psi_0     |\psi_1\ra =  0$, we combine these two systems by constructing the maximally-entangled $K+1$ qubit state 
\begin{equation}
|\Psi\ra =  \frac{  |0\ra |\psi_0\ra +  |1\ra |\psi_1\ra}{\sqrt{2}}  .
 \label{PSImax}
\end{equation}
The new auxiliary system has twice as many degrees of freedom\footnote{Technically it has \emph{more} than twice, because a $K$ qubit system has $2^{K+1} - 2$ real degrees of freedom, but this distinction is unimportant in the limit of large $K$.} as the  auxiliary system for  the original $K$ qubit quantum system. 
This is because the wavefunction has twice as many components. Thus we see that the addition of one qubit is the operation that doubles the auxiliary system.

If the states  $|\psi_0\ra$  and   $|\psi_1\ra$ are independently picked at random their relative complexity will almost always be maximal. In that case it can be shown that the complexity of $|\Psi\ra$ will be twice the complexity of either $|\psi_0\ra$  or   $|\psi_1\ra.$  

Let's suppose that the new qubit, which we'll call $\tau$, is uncoupled from the other qubits and that 
$|\psi_0\ra$ and $|\psi_1\ra$ are separately maximally complex. Thus the overall auxiliary system is two copies, each in complexity equilibrium.

Next we turn on generic $k$-local interactions between $\tau$ and all the other qubits. The overall system will come to complexity equilibrium with complexity
\beq
\CC_\textrm{final}= 2^{K+1} = 2^K + 2^K.
\ee
In other words the final complexity will be the same as the sum of the complexities of $|\psi_1\ra$ and $|\psi_0\ra.$ This is exactly like mixing two uncorrelated  gases of classically identical particles, each initially in equilibrium at the same temperature. The final entropy is the sum of the initial entropies and the process is reversible.

In the thermodynamic case it is obvious that no useful work can be extracted from such a process. In the complexity case, at all stages of the process the system is in a state of maximal complexity; thus according to our earlier discussion, no useful directed computational work can be done.

Now let's consider the case $|\psi_0\ra = |\psi_1\ra$. In this case, the extra qubit is not entangled with the rest of the system, which we continue to assume is in complexity equilibrium, 
\begin{equation}
|\Psi\ra =  \frac{  |0\ra +  |1\ra}{\sqrt{2}} \otimes  | \psi_0\ra.
\end{equation}
This time the
two auxiliary systems are in exactly the same state.

\begin{figure}[H]
\begin{center}
\includegraphics[scale=.2]{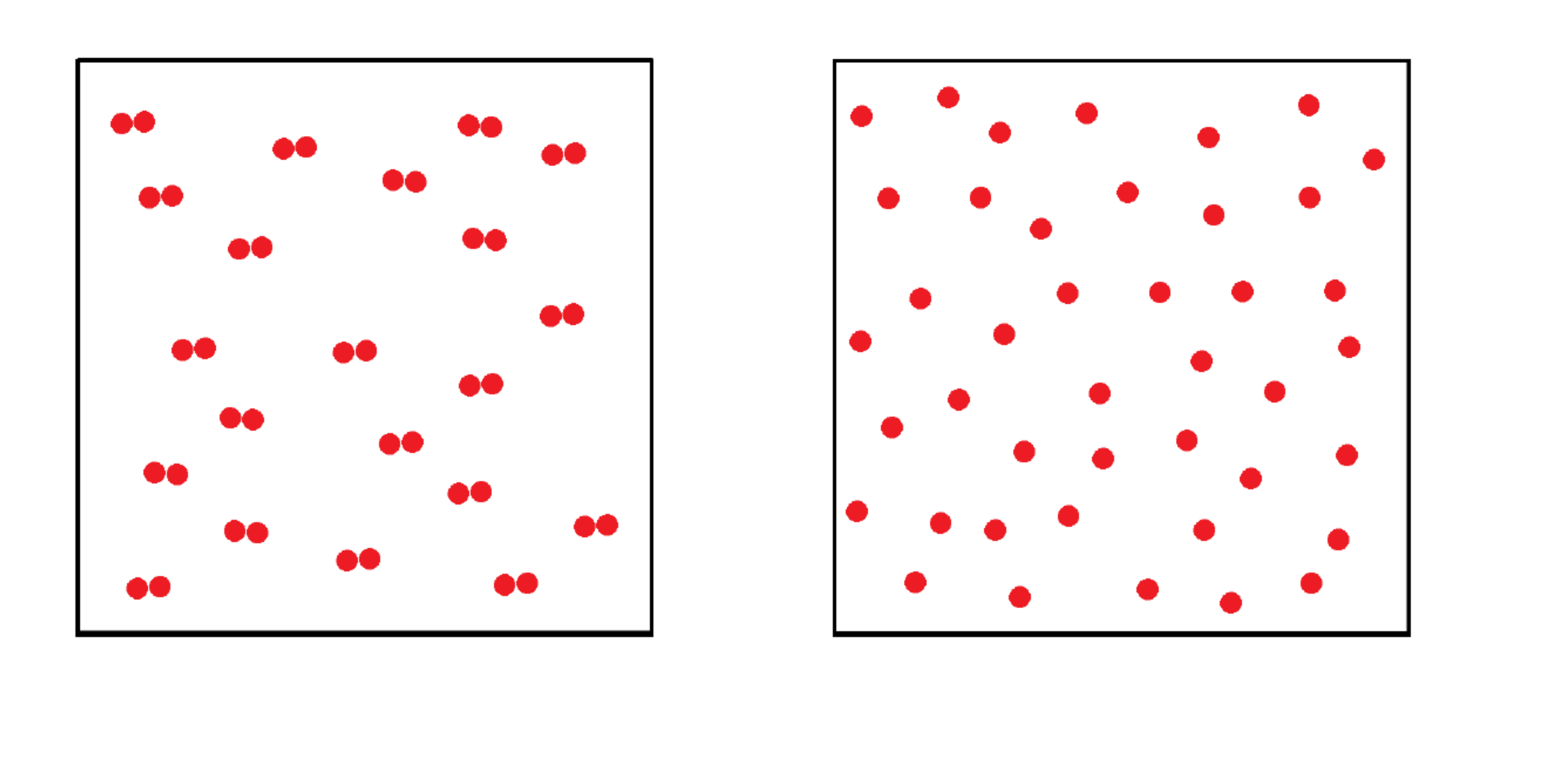}
\caption{The left panel shows a gas of $2N$ classical particles created with the particles paired. The entropy is the same as a gas of $N$ particles. In the right panel the gas has come to equilibrium and the particles become randomly distributed.  The entropy in the right panel is twice the entropy in the left panel. No work can be extracted from a gas in equilibrium,  but the gas of paired particles is far from equilibrium and so can be used to do work.}
\label{f4}
\end{center}
\end{figure}

The 
 initial complexity is $2^K,$ but after turning on an interaction that depends on the extra qubit and waiting for a long time,  the final complexity is $2\times 2^K,$ i.e.  double the initial complexity\footnote{This effect was the basis for the claim in  \cite{Susskind:2015toa} that dropping an additional thermal photon into a black hole doubles the time that the horizon will be transparent (firewall-free).}. Is there a thermodynamic analog to this situation? Indeed there is. Imagine creating  the two gases in exactly the same micro-state. Such a distribution is far from equilibrium: every particle of one gas is constrained to have exactly the same position and momentum as the corresponding particle of the other gas. The total initial entropy is the same as the entropy of one copy. However, perturbing one of the copies of the system, and then letting the whole system interact and come to equilibrium will result in a final entropy that is twice the initial. This is schematically illustrated in Fig.~\ref{f4}.
\bn

For a genuine  classical  system it  follows from the laws of thermodynamics that work can be extracted from the initial out-of-equilibrium state. In the quantum-complexity case this would correspond to a resource being available in a state of sub-maximal complexity. This resource---uncomplexity---can be used to do computational work.

\subsection{One Clean Qubit}\label{one clean qubit}

In this subsection, we will give an example of how uncomplexity can be used to do `computational work'. We will see that in the process, the uncomplexity is expended.

First consider a system that has no uncomplexity---a state of maximal complexity. A maximally complex state is very much like a maximally mixed density matrix as long as we restrict ourselves to reasonably simple experiments. If we act on such a state with a polynomial size circuit the complexity can only be reduced by a negligible fraction. For any measurement of a non-exponentially complex observable, the result will be Haar random, so again no useful computation can result from an initial maximally complex state. A quantum computer that runs for an exponential time and reaches maximal complexity becomes useless for computation.

Now consider adding to this maximally complex state a single additional qubit in a pure state. This doesn't change the complexity, but the maximal complexity doubles, so the complexity is now only half the maximal value. From the analogy with the two component out-of-equilibrium gas in Fig.~\ref{f4}, we should expect that the additional qubit, which has replenished\footnote{This exponential rejuvenation may have a remarkable consequence for black holes. In \cite{Susskind:2015toa} it is argued that black hole event horizons are only transparent if complexity is increasing, and so a black hole in complexity equilibrium would not have a transparent horizon. But as pointed out in that paper, throwing a single qubit in a pure state into an old maximally complex black hole rejuvenates the horizon for an additional exponential time.}  the uncomplexity resource, will allow us to once again perform useful computational work.

Computation that makes use of either a maximally mixed state (or a maximally complex state) plus just one additional unentangled qubit is called ``One Clean Qubit" computation.  Just how much power one clean qubit computation provides and how to quantify it is not certain but it is known to be able to efficiently solve problems including some classically hard problems  \cite{Knill:1998wi}. Known examples include calculating the trace of a unitary operator and estimating certain properties of Jones polynomials.  We'll review the illuminating example of calculating the trace of a unitary operator, which was first worked out in \cite{Knill:1998wi}. 

\bigskip

We suppose we have a unitary operator  $G$ in the space \sk. The operator $G$ is constructed as a known  product of a polynomial number of gates $G = g_N g_{N-1}....g_1.$ The goal is to approximate its trace. For simplicity let's only worry about the real part of the trace.

Begin with the space of states $CP(2^K-1)$. We will try to construct a $K$ qubit circuit such that a measurement of $\sigma_1^z$ will give some non-trivial information about the value of $\Tr G^{\dag} + \Tr G.$ Assume the circuit is initialized to the simple state $|00000...0\ra$.

\begin{figure}[H]
\begin{center}
\includegraphics[scale=.2]{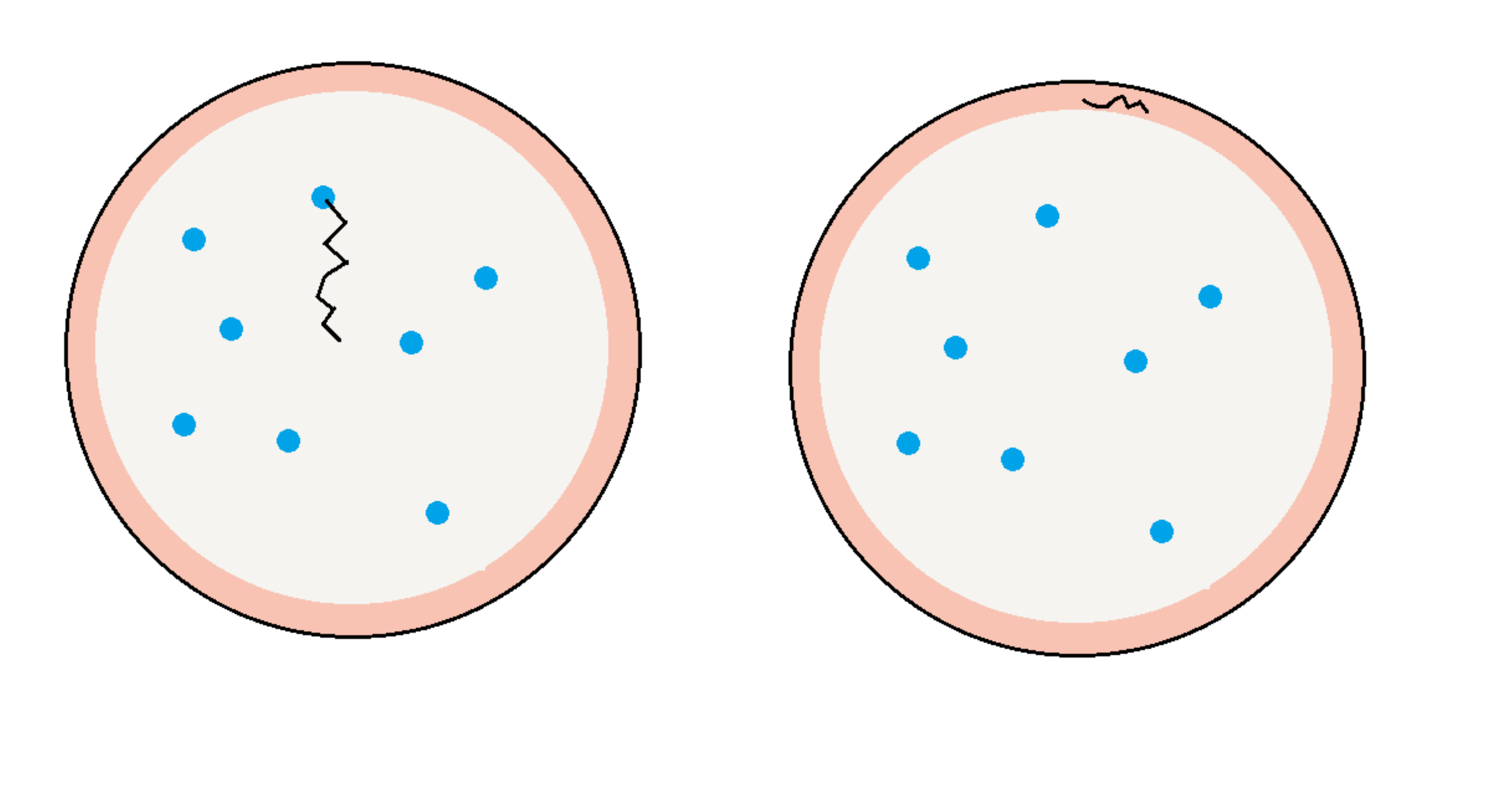}
\caption{The interior of the circles represent the space $CP(2^K-1)$ with the center point being the state $|0000\rangle.$ The blue regions are the target set. The left panel shows the evolution of a circuit programmed to get to a point on the target set. The trajectory is built from gates and each step increases the complexity.}
\label{f5}
\end{center}
\end{figure}

 Consider the neighborhood of all the states $|\psi\ra$ for which 
\be 
\la \psi| \sigma_1^z|\psi\ra = \Tr G \, .
\ee 
Call that the target set. If by running the circuit we can navigate to one of these points, then by a subsequent  measurement of $\sigma_1^z$ we learn something about $\Tr G^{\dag} + \Tr G.$   By repeating the experiment we can improve our knowledge. Thus the goal of directed computation is to decrease the relative complexity to zero  between the initial state and some  state that's in the target  set. Figure~\ref{f5} schematically illustrates the idea. The circles represent $CP(2^K-1)$ in a way such that distance from the center represents state complexity. In order to have a high probability of success it is important that each step increases the complexity.

\bn

Now let's suppose that instead of starting with the minimally complex state $|000..00\ra$ we start with a state in the darker pink outer region where the complexity is maximal $\sim 2^K.$ There are no blue points in this region since the expectation value of any observable is Haar random.
With overwhelming probability any gate that acts on a state in the dark pink region will leave the point in that region. This shows that directed computation is not possible starting with a state of maximal complexity, i.e., $\Delta \CC = 0.$

\begin{figure}[H]
\begin{center}
\includegraphics[scale=.2]{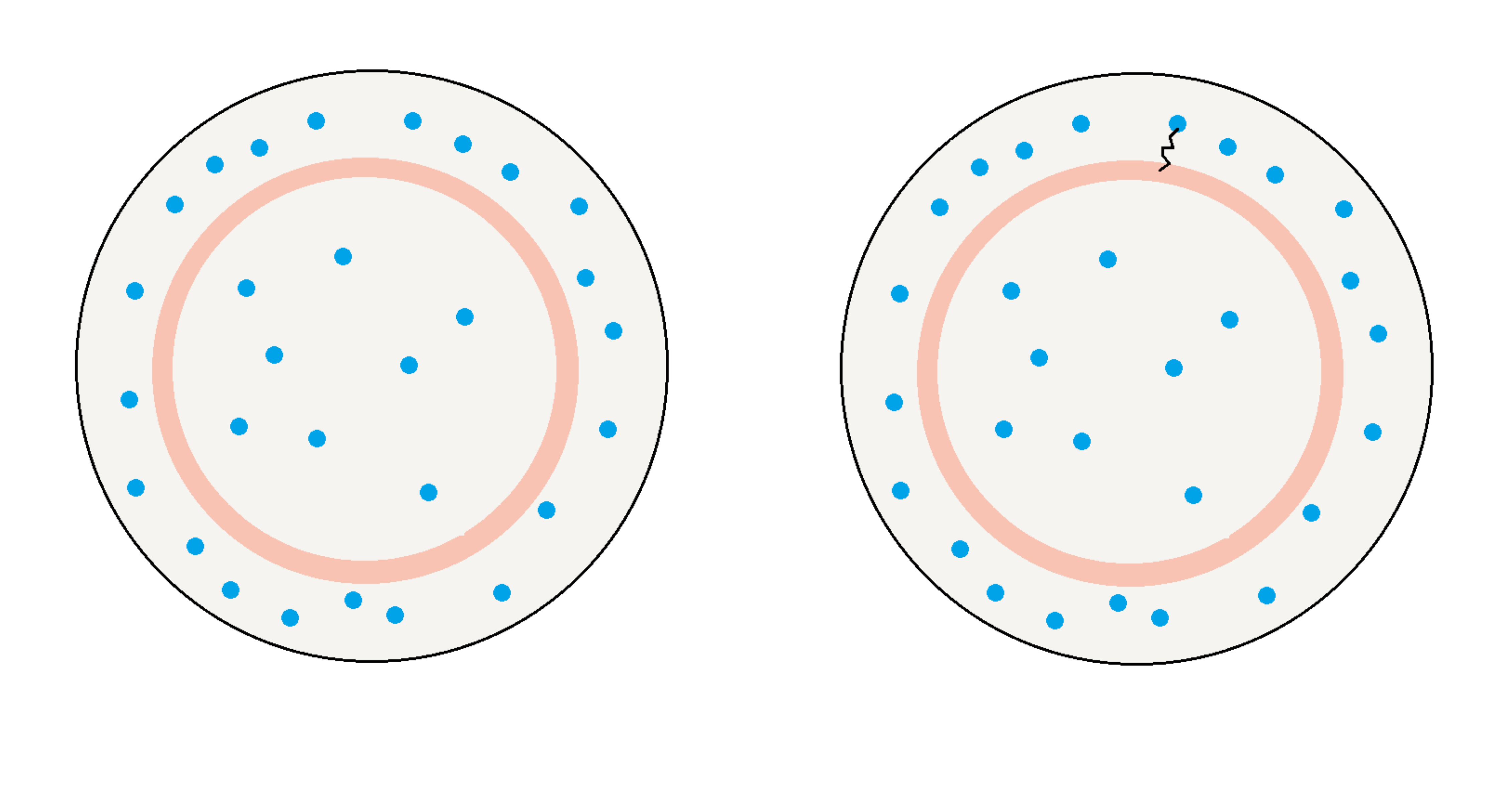}
\caption{Left: adding an extra qubit doubles the maximum complexity (adding an annulus to the space of possible states) and replenishes the uncomplexity resource. Right: given the additional resource, what was previously a state of maximal complexity now has some uncomplexity and can be used to further computation; this is illustrated by showing how a target state can be reached from the original maximally complex configuration.}
\label{f6}
\end{center}
\end{figure}

But now let us add one clean qubit $\tau$, thereby doubling the maximal complexity. The larger circles in Fig.~\ref{f6} represent $CP(2^{K+1}-1),$ the space of $K+1$ qubit states. The darker pink still shows states of complexity $2^K,$ but the region beyond it goes out to twice that complexity.

\bn

Note that the initial state for the one-clean-qubit calculation is in the dark pink region, but now we can reach blue dots by moving outward towards increased complexity; we don't have to fight against the second law.

The actual algorithm is simple \cite{Knill:1998wi} and we will describe it now. 
\begin{figure}[H]
\begin{center}
\includegraphics[scale=1.15]{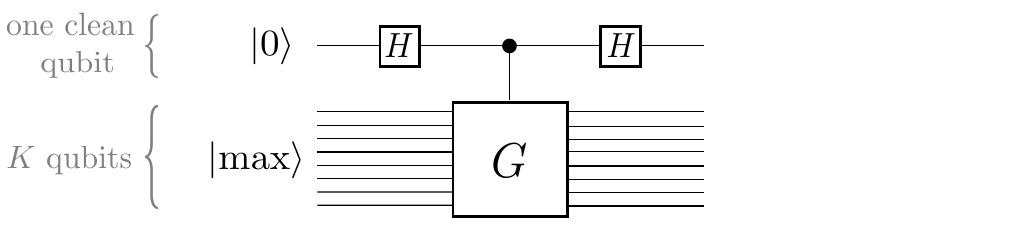}
\caption{Circuit for using a clean qubit to compute $\Tr [G].$ The input state has a total of $K+1$ qubits---the one clean qubit (top) and the $K$ maximally complex qubits (bottom). The symbol $H$ represents a Hadamard gate acting on the clean qubit. The clean qubit acts as a control for the circuit $G$: the circuit applies $G$ to the other $K$ qubits if the clean qubit is $|1\ra$, and does nothing if the clean qubit is $|0\ra$.}
\label{C1}
\end{center}
\end{figure}
Consider the quantum circuit shown in Fig.~\ref{C1}. The initial state is
\be 
|0\ra \otimes |{\rm{max}}\ra,
\ee
where $|{\rm{max}}\ra$ is any state of the $K$ qubit system with maximal complexity. We act with the first Hadamard gate\footnote{The Hadamard gate is defined by the matrix $(\tau^z + \tau^x)/\sqrt{2}$.} to give
\be 
\frac{|0\ra + |1\ra}{\sqrt{2}} \otimes |{\rm{max}}\ra.
\ee
Next apply the controlled $G$ operation $G_c = |1 \ra \la 1|  G + |0 \ra \la 0 | \mathds{1}$, where $G = g_N g_{N-1}....g_1$; this circuit applies $G$ to $|{\rm{max}}\ra$ if the control qubit is $|1\ra$, and otherwise leaves it unchanged. This gives 
\be 
 \frac{ |1\ra \otimes G |\rm{max}\ra  +|0\ra \otimes  |\rm{max}  \ra  }{\sqrt{2}}.
\ee
Now acting with the second Hadamard yields
\be 
|0\ra \otimes \frac{1+G}{\sqrt{2}}|\rm{max}\ra +|1\ra \otimes  \frac{1-G}{\sqrt{2}}|\rm{max}\ra.
\ee
This completes the computation. To make use of it we note that the
 expectation value of $\tau^z$ is given by
\be 
\la \tau^z \ra =  \la {\rm{max}}| G^{\dag}+G |{\rm{max}}\ra.
\ee
This in itself is not useful for our purpose---determining $\Tr G$---but because
 $|{\rm{max}}\ra$ is a maximally complex $K$ qubit state, with overwhelming likelihood 
\be 
\la {\rm{max}} |G| {\rm{max}}\ra = \Tr G. 
\ee
Thus by applying the circuit $HG_c H$ we have set up a state in which we can learn something about $\Tr G$ by making a measurement of $\tau^z.$

The measurement itself cannot be represented as an operation in the classical auxiliary system. As we said earlier it should not be considered as part of the computational work. The computational work is associated with the process that went into setting up the state, i.e., acting with the circuit $HG_cH$,  and only then at the very end do we allow a  measurement. 
By repeating this experiment, including the measurement, over again with fresh clean qubits we can get an arbitrarily accurate estimate for $\Tr G$.
 
 \bn
 
In classical thermodynamics we can repeat an operation designed to raise a weight one meter and thereby raise it two meters, three meters, four meters, and so on until we run out of resource.
 The same is true of computational work. For example by repeating the circuit of Eq.~\ref{circuit} in the form
\be 
\Big(H G_c H\Big)^n  |0\ra \otimes |\rm{max}\ra 
\label{circuit}
\ee
one can determine information about the trace of $G^n.$ (As before we only make a measurement at the end.)  For obvious reasons the problem of determining the trace of a higher power of $G$   becomes more difficult as the power increases. It is also clear that the repeated action of the circuit depletes the resource, roughly by the complexity of $G$ each time it is repeated.

`One clean qubit' computation is an example of using uncomplexity to do computational work. It exhibits all three of the criteria that we listed at the start of Sec.~\ref{Sec: combining}. 
\begin{enumerate}
\item First, it implements a transition that  is directed towards a goal---the goal of calculating the trace of $G$.
\item  Second, it uses up a resource---at the end of the computation, the additional qubit is no longer clean, and the complexity of the $K+1$ qubit state has increased by approximately the complexity of $G$. Or to put it another way the uncomplexity resource has diminished by that amount. 
\item  Third, the process involves a transition from one macroscopic state to another by a procedure that does not depend on the microscopic state---we extracted information about $\Tr G$ without knowing precisely which state we started or ended in. (Thus no Maxwell's Demons were involved. Instead we did something analogous to doing work by expanding the volume of a gas using a procedure that does not require knowledge of the starting microstate.) 
\end{enumerate}

It would be very interesting to know how the power of one clean qubit is connected to the doubling of the maximum complexity, and whether it is similar to the ability to do work with a system of identical gases in which the molecules are paired in a non-thermal distribution.

\subsection{Kolmogorov Uncomplexity as a Resource} \label{subsec:Kuncomplexity}

We have argued that computational `uncomplexity' is a resource that can be used to do directed quantum computation. But computational complexity is not the only kind of complexity that has arisen in this paper. In Sec.~\ref{KC_and_KE} we argued that while the \emph{positional} entropy of $\CA$ corresponds to the \emph{computational} complexity of $\CQ$, the \emph{kinetic} entropy of $\CA$ corresponds to the \emph{Kolmogorov} complexity. This therefore raises the question of whether Kolmogorov uncomplexity is also a resource. 

The answer is yes, but we will see that the resource is useful for a rather different purpose than computational uncomplexity. This means that from the point of view of the $\CQ$-$\CA$ correspondence this subsection is something of an aside, but it is well worth explaining. In this subsection we will explain that Kolmogorov uncomplexity is a resource that is useful for doing erasure.

This is beautifully illustrated by an example in an old paper of Bennett, Gacs, Li, Vitanyi, and Zurek  \cite{Bennett1993}, which examines apparent violations of Landauer's Principle \cite{Landauer}. Landauer's Principle says that, while reversible transformations can be performed without free-energy cost, to erase a bit (to reset it to zero no matter whether it starts at one or at zero) requires a free energy of $k_B T \log 2$. However, there are some examples where bits can seemingly be reset much more cheaply than this. 

What Bennett et al.~show is that these apparent violations occur precisely in cases that have Kolmogorov uncomplexity, since in those cases the states can be compressed before being erased. (For example, it requires less free energy to erase the first million digits of $\pi$ than to erase a million random digits. This is because it is possible to reversibly transform the first million digits of $\pi$ to the much shorter computer program that outputs them. Since this compressed description has much less Kolmogorov uncomplexity than the original description,  performing the compression expends uncomplexity.) Specifically, they show that the free energy cost of deleting a bit string is not given by the total number of bits, but by the Kolmogorov complexity of the bit string. For generic bit strings these two coincide, but for special low complexity strings the Kolmogorov complexity is less. The total saving compared to a naive application of Landauer's principle is given exactly by the uncomplexity, 
\begin{equation}
\Delta F \Bigl|_\textrm{saved}   = k_B T \log 2 \, \, \Delta C_{\kappa}.
\end{equation}
The Kolmogorov uncomplexity of one bit string can be used to erase another bit string; in the process, the resource is expended. \\

\noindent We thus see that both \emph{computational} uncomplexity and \emph{Kolmogorov} uncomplexity can be used to carry out information theoretic tasks.

\sc
\section{Uncomplexity as Spacetime} \label{sec:uncomplexityasspacetime}
  
Our original interest in complexity theory began with the question: How does one describe the interior of a black hole in holographic terms? In this section we would like to come back to that question in light of the conjecture that uncomplexity is a resource. We will see that the \it black-hole/complexity-connection \rm  provides a  new way to think about uncomplexity as a spacetime resource\footnote{We would like to thank Douglas Stanford for a critical remark that led to this section.} based on classical general relativity (GR). In particular classical GR provides another way to think about the rejuvenating power of one clean qubit. 

To understand the uncomplexity resource in GR terms let's suppose  
  Alice is a black hole explorer stationed just outside a one-sided AdS  black hole at boundary time $t$. She intends to jump from the AdS boundary into the black hole. The resource that she cares about is spacetime volume---without which she will perish at the horizon.

\begin{figure}[H]
\begin{center}
\includegraphics[scale=.2]{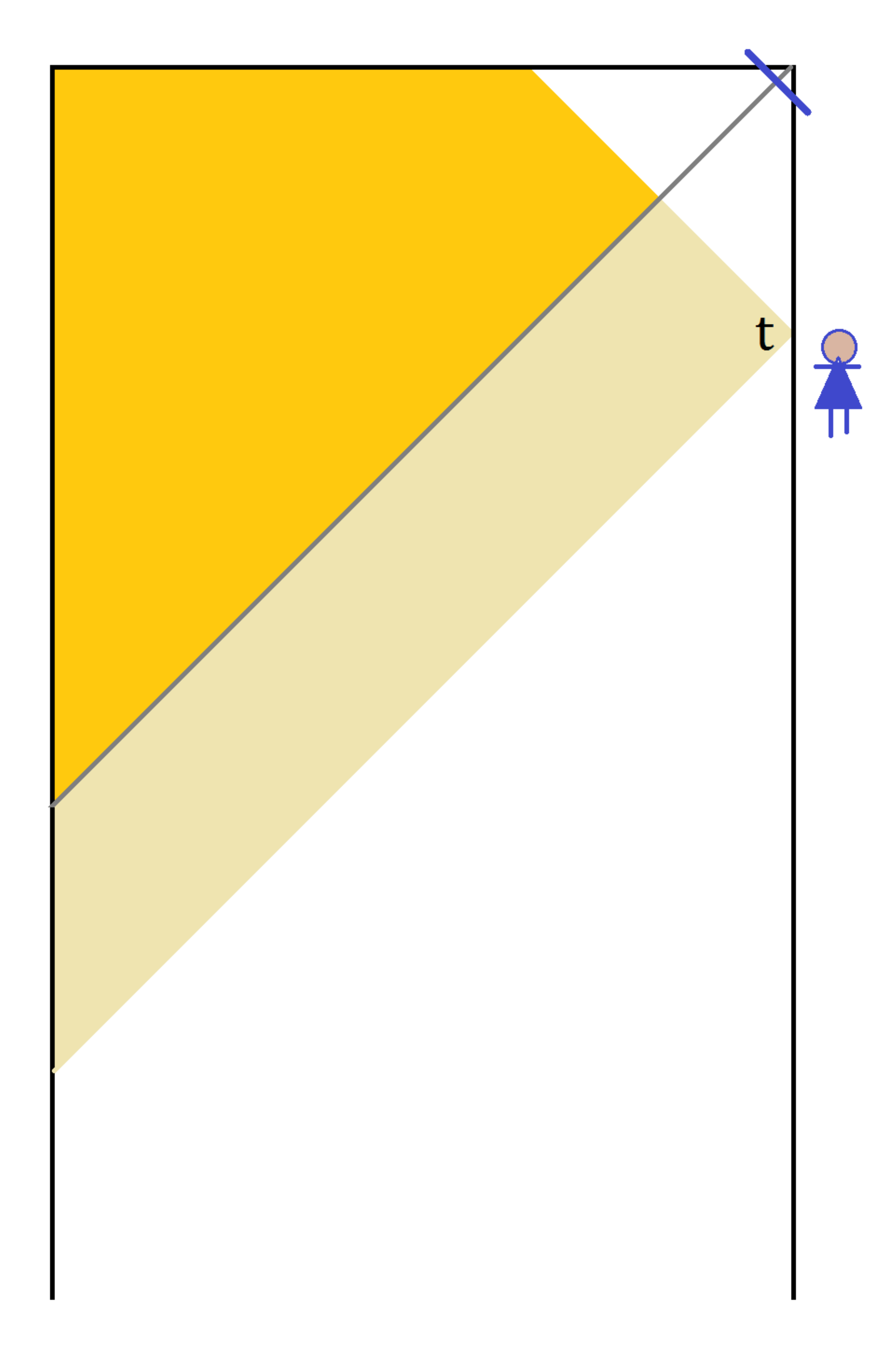}
\caption{The Penrose diagram for a one-sided black hole in AdS. The WDW patch anchored at a boundary time $t$ is shaded yellow; the part of it behind the event horizon is dark yellow. The blue line in the top right is a null geodesic emanating from the point at which the boundary state complexity becomes maximal.}
\label{r1}
\end{center}
\end{figure}

Recall that the quantum state of the black hole interior has a growing complexity (for $t>0$) that is dual to the growing spacetime volume behind the horizon. At any instant the complexity is given by the Einstein-Hilbert action of the Wheeler-DeWitt  (WDW) patch anchored at time $t$  on the boundary  \cite{Brown:2015bva}\cite{Brown:2015lvg}. The part of the WDW patch outside the horizon has a time-independent divergence, which after initial transients can be regulated by considering only the portion of the space behind the horizon, as shown in Fig.~\ref{r1}. The action of the dark yellow region behind the horizon is of order its spacetime volume. 

Slightly simplifying the discussion, we can say that the complexity is given by the spacetime volume $V_4$ of the dark yellow region, multiplied by some numerical factors that depend on the gravitational constant $G$ and the AdS radius of curvature $\ell_\textrm{AdS}$,
\be 
\CC(t) \sim \frac{V_4 }{G \ell_\textrm{AdS}^2}.
\ee

A straightforward GR calculation shows that the action increases linearly with time, with a coefficient equal to the mass of the black hole. This is consistent with the early growth of complexity in Fig.~\ref{f1}. It is believed that the classical description of the black hole breaks down when the complexity stops increasing, once $\CC = \CC_\textrm{max}.$ This occurs at $t_\textrm{max}=e^S$ at which time the horizon becomes opaque by developing a firewall \cite{Susskind:2015toa,Almheiri:2012rt}. 
In Fig.~\ref{r1} the cutoff at $t_\textrm{max}$ is shown as a blue diagonal slash in the upper right corner of the diagram. Time does not literally run out at the cutoff, but because complexity is bounded by $\CC_\textrm{max}$ the classical growth of the black hole interior must break down.

\begin{figure}[H]
\begin{center}
\includegraphics[scale=.2]{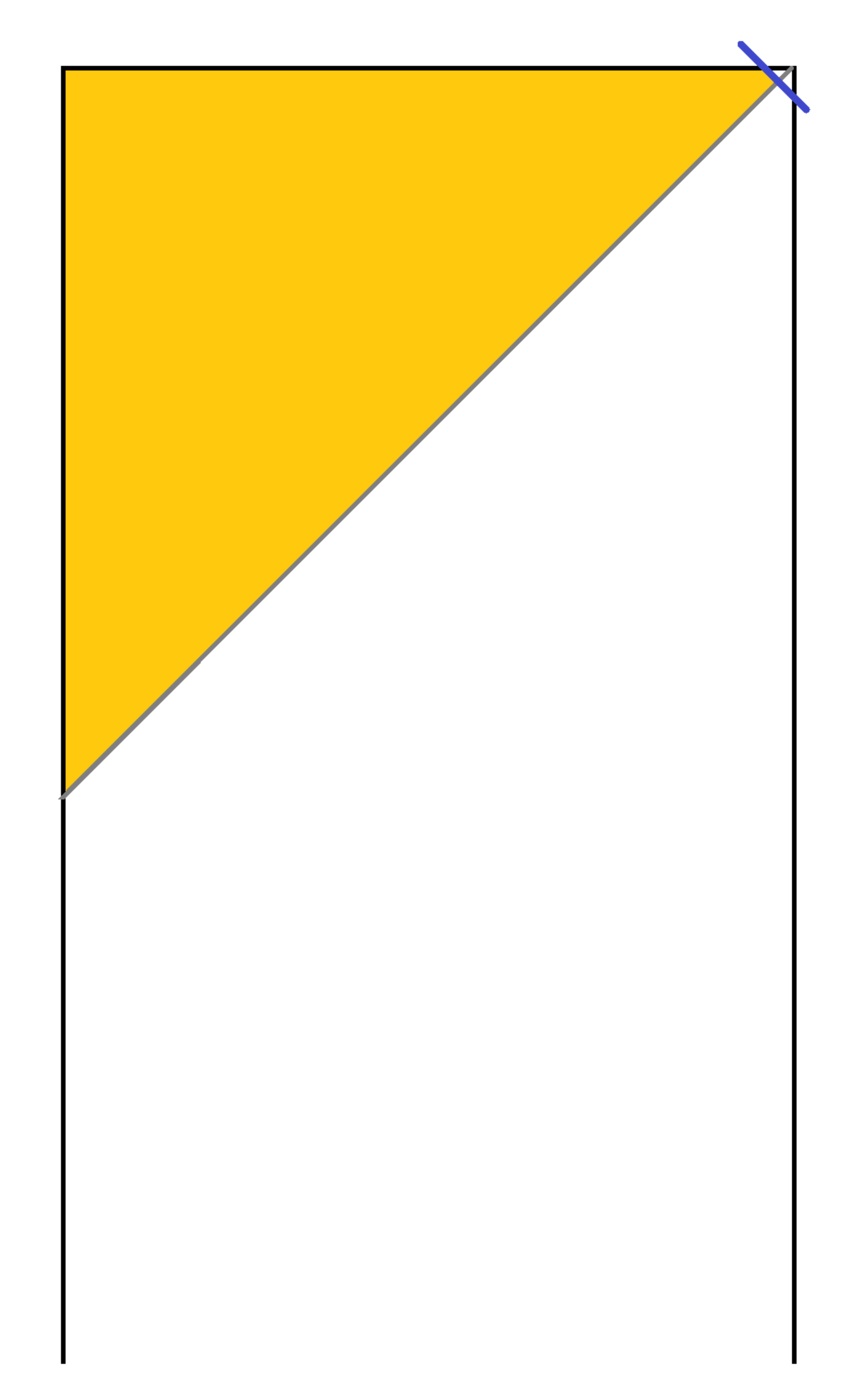}
\caption{The WDW patch for maximum complexity.}
\label{r2}
\end{center}
\end{figure}

Let us consider in more detail  the maximum complexity. Figure \ref{r2} shows the WDW patch pushed up to the cutoff time. The maximum complexity $\CC_\textrm{max}$ is the action of this new WDW patch.  Classically the action (4-volume) in the upper corners would grow indefinitely, but the cutoff at $t_\textrm{max} \sim e^S$ keeps it finite.

\begin{figure}[H]
\begin{center}
\includegraphics[scale=.2]{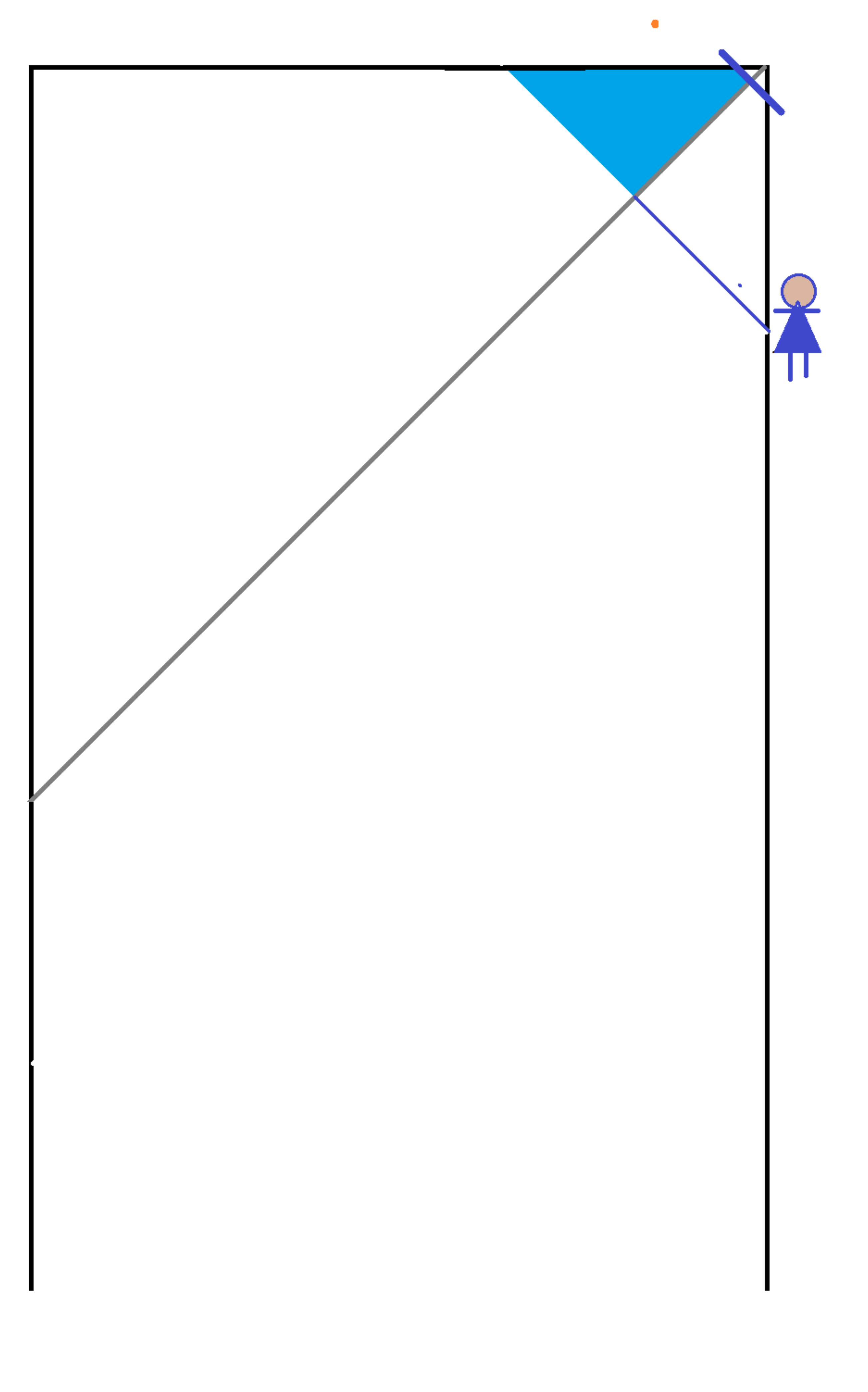}
\caption{The uncomplexity is proportional to the spacetime volume of the blue region.}
\label{r3}
\end{center}
\end{figure}

The uncomplexity $\Delta\CC(t) \equiv \CC_\textrm{max} - \CC(t)$ is  given by the action of the dark yellow region of Fig.~\ref{r2} minus the action of the dark yellow region of Fig.~\ref{r1}. This difference is shown in blue in Fig.~\ref{r3}. 

\bn
The uncomplexity is proportional to the 4-volume of the  blue triangular region, which is cutoff at $t_\textrm{max} \sim e^S$. This 4-volume is finite, and goes to zero as $t \to e^S.$ 

We  see something interesting from the figure.  The blue region  may be identified with the  union of all interior  locations behind the horizon that Alice can visit  if she enters the black hole at any time after $t.$
The uncomplexity therefore represents the spacetime resource available to an observer who intends to enter the horizon.

\bn

Suppose Alice wishes to jump in after the black hole has become maximally complex. According to \cite{Susskind:2015toa}  she will run into an obstruction at the horizon. The situation is analogous to attempting to compute with a computer that has reached maximal complexity; the resource will have been exhausted. Can Alice do anything to renew the resource? 

\begin{figure}[H]
\begin{center}
\includegraphics[scale=.2]{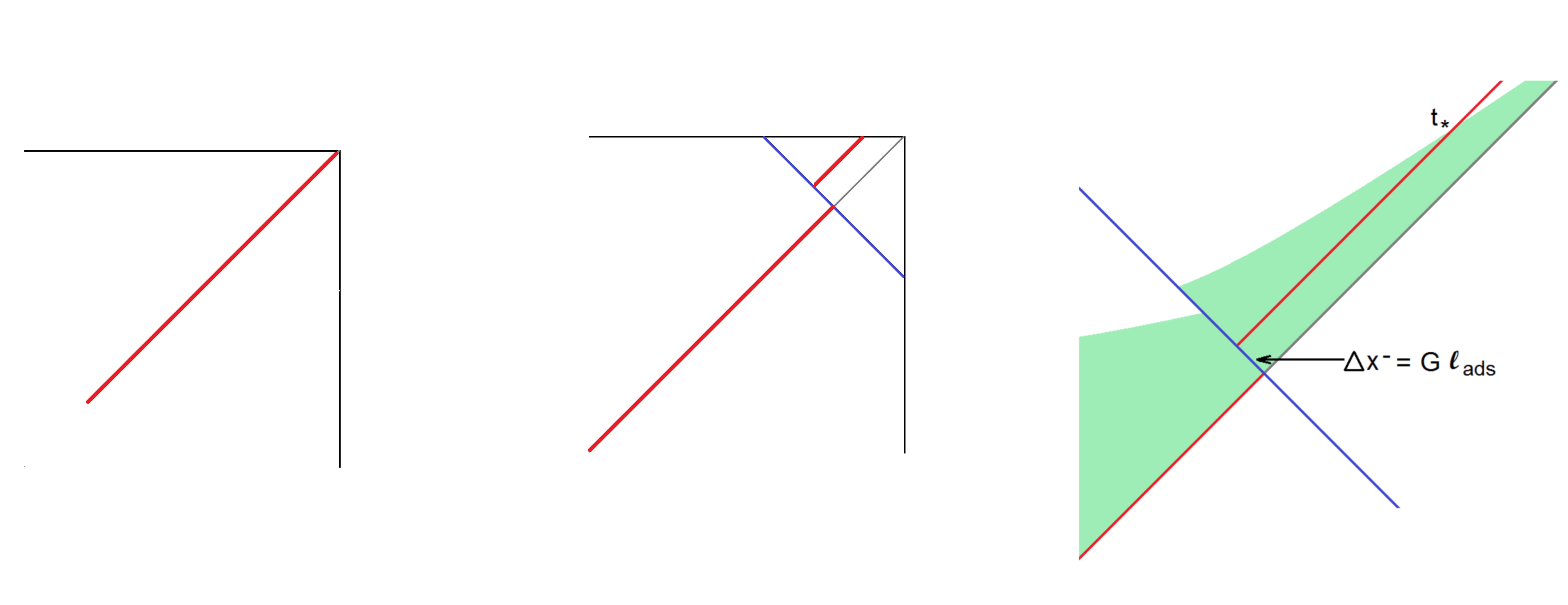}
\caption{The first panel shows the upper-right corner of the black hole Penrose diagram with the red line representing an opaque horizon that would be expected for a black hole of maximal complexity. The opaque horizon can be modeled by an infinitely thin Shenker-Stanford gravitational shockwave. The blue line in the second panel is a thermal  quantum injected from the boundary. Such a quantum increases the entropy of the black hole by one bit. The effect of the low energy quantum is to shift the shockwave up and to the left thus separating it from the horizon. In a scrambling time it will be lost into the singularity. The right panel which was taken from \cite{Susskind:2015toa} shows the process in more detail. The upshot is that within a scrambling time the horizon has become transparent; this newfound transparency lasts for an exponential time.}
\label{s4}
\end{center}
\end{figure}

As explained in  \cite{Susskind:2015toa}, all Alice has to do is to throw in one thermal photon and wait a scrambling time. This will restore the transparency of the horizon for an additional exponential time, in the same way that the computing power of a maximally complex computer can be restored by adding a single clean qubit. This phenomenon is essentially a classical GR effect, which we illustrate in Figure \ref{s4}.

The conclusion drawn in \cite{Susskind:2015toa} is that the obstruction at the horizon due to maximal complexity will be removed by adding to the black hole one clean qubit in the form of a thermal quantum.
The rejuvenating effect of the added qubit parallels the effect in a quantum computer that has reached maximal complexity.

\sc
\section{Summary}

Let us summarize the material in this paper:

\bi 
\item  Section~\ref{sec: Qsystem} introduced the class of quantum systems $\CQ$ that we study, namely \kl \ systems composed of $K$ qubits interacting through a Hamiltonian which is a sum of terms, each containing no more than $k$ qubits. Alternatively the qubits may interact in a \kl \ quantum circuit built of gates with no more than $k$ qubits. Such systems are typically fast scramblers.

 We explained that the evolution of complexity for a \kl \  system of $K$ qubits closely resembles the classical evolution of entropy for a system of $\exp[K]$ classical degrees of freedom and raised the question of the source of this similarity.
 
We also explained the SYK strategy of averaging over randomly chosen time-independent Hamiltonians. This sometimes allows us to determine the average behaviors  for problems which are too difficult to solve in individual instances.

\item In Sec.~\ref{sec: classical system} we formulated the evolution of the time-development operator $e^{-iHt}$ as a classical mechanics problem of an ``auxiliary" system $\CA.$ The system $\CA$ consists of 
a non-relativistic particle moving on the space \sk. The  auxiliary system  for a system of $K$ qubits has  a number of classical degrees of freedom exponential in $K$.

The first-order Schrodinger equation of $\CQ$ is replaced by a second-order equation of motion for $\CA$, in which the Hamiltonian is eliminated altogether, in favor of initial conditions on the velocity of the particle. Averaging over a Maxwell-Boltzmann ensemble of initial velocities is equivalent to averaging over quantum Hamiltonians as in SYK.

\item  The usual inner-product metric on either the space of states or the space of unitary operators is poorly suited to studies of quantum chaos. Section~\ref{sec:geo-comp} is devoted to the concept of relative complexity: a metric which represents the degree of difficulty in making a transition between two states, and also of doing an interference experiment that measures the relative phase between states. Relative complexity can also be defined for unitary operators and has a similar meaning.

The ``complexity metric" defined by relative complexity is much better suited to measuring the difference between states of a chaotic system than the standard inner product metric. Inspired by the work of Nielsen and collaborators \cite{Nielsen,Dowling}, Sec.~\ref{sec:geo-comp} works out the basic mathematical properties of complexity metrics and shows that they are closely related to the negatively curved geometry of the toy model. In particular we calculated sectional curvatures and  showed behavior consistent with the toy model.

\item Section~\ref{sec: particle} introduces the $\CA$ system as a classical nonrelativistic particle moving on this complexity geometry. The relative complexity of two unitary operators is the minimal action required to go from one to the other subject to a constraint on the auxiliary energy of the particle. 

\item  Section~\ref{sec: stat-mech} introduces our basic conjecture relating classical entropy to quantum complexity. We argued that after averaging over Hamiltonians (as in SYK) the ensemble-average of quantum  complexity is equal to the classical entropy of the auxiliary system $\CA.$ In order to make this identification complete we must include not just the circuit complexity---the number of gates in the circuit---but also the Kolmogorov  complexity of the algorithm that the circuit implements. In the case of a Hamiltonian quantum system the Kolmogorov complexity is the length of the shortest program needed to specify the Hamiltonian. Unlike the gate complexity, it does not grow linearly with time and so soon becomes negligible compared to the gate complexity. 

The connection between quantum complexity and classical entropy is the link that suggests a thermodynamic description  of complexity. In Sec.~\ref{sec: second-law} we used this connection in order to formulate a Second Law of Complexity which is really just the second law of thermodynamics for the auxiliary classical system $\CA.$ This line of reasoning explains the observation in Sec.~\ref{Sec: QC and CE} that the evolution of complexity for a $K$ qubit system behaves like the evolution of entropy for a system with a number of classical degrees of freedom exponential in $K$.

\item In Sec.~\ref{Sec: combining} we discuss the concept of uncomplexity---the gap between the  complexity of a state and the maximum possible complexity---and give evidence that it is a resource  useful for doing computational work. An important component of  resource theory \cite{Spekkens} is combining systems into bigger systems. In the present framework this means combining auxiliary systems. Surprisingly, combining two auxiliary systems has nothing to do with combining the corresponding quantum systems. To double the size of an auxiliary system one only needs to add a single qubit to the quantum system. We illustrate the idea of uncomplexity as a resource with the example of ``one clean qubit" computation.

\item 
Finally, in Sec.~\ref{sec:uncomplexityasspacetime}, we look at the holographic dual to the uncomplexity of a boundary state.  We show that when a black hole is present the  resource---uncomplexity---is the total spacetime volume accessible to an observer who plans to cross the horizon.

\ei

\sc
\section{Questions}
The  strategy of averaging over an ensemble of Hamiltonians (in computer science this would amount to averaging over algorithms) may allow one to solve problems about average behaviors that would be much too hard for individual instances. We are raising the possibility that very difficult problems of complexity theory may be solved on average by classical statistical mechanics and thermodynamics. As an example we point to the correspondence between the  evolution of quantum complexity---an extremely difficult problem for specific Hamiltonians---and the classical evolution of entropy---a merely hard problem.

\bn

\noindent This paper raises many questions, a few of which we will mention here.

\bi
\item{Definition of Complexity} 

We have assumed that there is a robust concept of complexity, but in fact there is a large family of complexity measures. It is important to understand how they are related and whether a preferred measure of complexity can be identified. In the context of the complexity geometry the different measures are encoded in the moment of inertia tensor $\CI_{IJ}.$ We showed that the sectional curvatures will generically be negative and order $1/K$ (in agreement with the toy model) as long as the penalty factors are not too small. What are the rules governing the choice of $\CI$, how should its elements grow with increasing weight, and is the curvature approximately constant as predicted by the toy model of \cite{Brown:2016wib}?

\item{Counting}

The conjecture that average complexity and classical entropy are the same rests on the assumption that the number of unitary operators with complexity less than or equal to $\CC$ grows like $e^{\CC}$ for sub-maximal $\CC.$
We were able to give  arguments in the stochastic context and for state-complexity,  but the arguments are far from a proof. Proving the conjecture requires counting the unitaries on $2^K$-dimensional tori in \sk.

\item{Local vs Global Chaos}

The motion of the $\CA$-system with a time-independent Hamiltonian is generically ergodic.
Whether or not it is chaotic seems to depend on what metric we attach to \sk. According to the bi-invariant metric all sectional curvatures are positive which implies that geodesics converge.

On the other hand, as we emphasized in Sec.~\ref{distance}, conventional inner-product metrics do not capture an important concept of distance between states, or for that matter, between unitary operators. Distances in the bi-invariant metric are bounded by $\pi/2$ but complexity distances can grow to enormously large values. Evidently complexity distances between neighboring trajectories can grow exponentially with time whereas the inner product distances do not, although in both metrics the system is ergodic.

The question is whether the motion in complexity geometry is genuinely chaotic, and does it matter? True classical chaos is often diagnosed by the spectrum of Lyapunov exponents, with a single positive Lyapunov exponent indicating chaos. The concept of a Lyapunov exponent is a global one, defined by the infinite time average of trajectory deviation. By contrast there is also a concept of local Lyapunov exponents, which diagnoses local deviation, and local unpredictability. Local Lyapunov exponents are positive in regions of negative curvature.
When local Lyapunov exponents are positive the system will behave chaotically for a length of time, but over sufficiently long times it may only be ergodic. Of course if this time is long enough---say for example exponential in $K$, the distinction between global and local chaos may be unimportant.

Our guess is that the $\CA$-system (with the complexity metric) is locally  chaotic over an exponentially long time, but that it is not truly chaotic.  But by then it hardly matters.

\item Classical Complexity

In this paper we have been concerned with the thermodynamics of \emph{quantum} computational complexity. However, many of the arguments would apply to \emph{classical} computational complexity. Can we also define a thermodynamics of classical computational complexity?

\item{Least Action and Least Computation}   
 
We can ask about the action-complexity connection. By now we have several versions of \it Action Equals Complexity\rm. In \cite{Brown:2015lvg} it was conjectured that the principle of least action for a gravitational system might ultimately become a principle of least computation. In this paper we have proposed another least action principle for the auxiliary system $\CA,$ which would also describe the evolution of the state of a black hole. The question is: what is the relation between these apparently different but similar principles of least action/computation? More specifically, are they somehow the same? A similar suggestion in a slightly different context was recently proposed in \cite{Chemissany:2016qqq}.

\item{Uncomplexity as a Resource}

One of the most interesting questions raised by this paper is whether there is a sense in which the gap between quantum complexity and maximal quantum complexity---the uncomplexity---is a quantitive measure of a resource useful for quantum computation.  Can we precisely characterize the resource and does it fit into standard resource theory \cite{Spekkens}? 

Can we understand the interplay between computational uncomplexity and Kolmogorov uncomplexity? 

\item{First Law of Complexity}

In this paper we have argued for the existence of a second law of complexity. Identifying a first law of complexity is left for future investigation.

  The conventional theory of thermodynamics was developed through a sequence of thought experiments involving adiabatic compression, heat engines, refrigerators, and the vanquishing of Maxwell's demon. Can we come up with a set of parallel thought experiments involving quantum complexity? What will be the steam engine of quantum computation?

\ei
\bn

\section*{Acknowledgements}

We'd like to thank Patrick Hayden and Scott Aaronson for patient tutoring and especially for introducing us to the idea of one-clean-qubit computation. We'd also like to thank Andy Albrecht,  Beni Yoshida, Dan Roberts, Daniel Harlow, Douglas Stanford, Henry Lin, Hrant Gharibyan, Nicole Yunger Halpern, Simeon Hellerman, and Ying Zhao.  This publication was made possible in part through the support of a grant from the John Templeton Foundation; the opinions expressed in this publication are those of the authors and do not necessarily reflect the views of the John Templeton Foundation.
\sc
\appendix

\section{Some Terminology} \label{sec:terminology}

For easy referencing, we list here some terminology and notations.  

\bn
1) The two-dimensional model of \cite{Brown:2016wib} will be referred to as the toy  model. The toy model represents quantum evolution as the motion of a non-relativistic particle on a two-dimensional hyperbolic space with a curvature of order $1/K.$

\bn
2) The space of special unitary operators acting on $K$ qubits  (or $2K$ real fermion operators) is \sk. Elements of \sk  \ are denoted $U, \ V, \ W,....$ The Pauli basis for the generators of $\suk$ consist of: the  Pauli operators
$\sigma^a_i,$ where $a$ labels the three axes $x,y,z$ and $i$ labels the $K$ qubits; and all products of Pauli operators for multiple qubits. In all there are $(4^K-1 )$ such generators. They will be labeled $\sigma_I$ where $I$ runs over  $(4^K-1 )$ values.

The weight of a $\sigma_I$ is the number of single qubit Pauli operators that it contains. Thus for example the weight of $\sigma^x_1$ is $1,$ and the weight of $\sigma^x_1\sigma^y_3\sigma^y_4$ is $3.$

\bn
3) $J_I$ is a coefficient or coupling constant in the quantum Hamiltonian of the $\CQ$-system.

\bn
4)  The conventional bi-invariant metric on \sk \ is called the standard metric. (This is different than Nielsen's usage.) Bi-invariant means that the metric is invariant under left and right-multiplication by unitary operators. The standard metric can be written,
\be 
dl^2 = \Tr [ dU^{\dag} \ dU ] .
\label{symmetric}
\ee
(The $\Tr$ notation denotes normalized trace, i.e., $\Tr {\mathds{1}} = 1.$)

\bn
5) The complexity metric  \cite{Nielsen}\cite{Dowling} is right-invariant but not left-invariant.  The precise form of the metric is given in Eqs.~\ref{right-inv} \& \ref{Omega}.

\bn
6) The complexity metric is written
\be
dl^2 = G_{MN} dX^M dX^N,
\ee
where the $X's$ are coordinates on \sk.

\bn
7) The classical auxiliary system defined in  Sec.~\ref{sec: classical system} is denoted $ \CA.$ The original quantum system of $K$ qubits with Hamiltonian given by Eq.~\ref{eq:$k$-local} is denoted $\CQ.$

\bn
8) A subscript $a$ indicates that a quantity refers to the auxiliary system, not the quantum system. Thus
 $V_a$ represents the magnitude of the velocity of the particle in the auxiliary system. $E_a$ indicates the energy of the $\CA$ system, etc.

\bn
9) The time $t$ is measured in dimensionless units. For an uncharged neutral black hole the time $t$ differs from the asymptotic  Schwarzschild  time $t_{schw}$ by a factor $\beta/2\pi,$
\be 
t=\frac{2\pi }{ \beta} t_{schw}, \label{eq:tintermsoftschw}
\ee
where $\beta$ is the inverse temperature of the black hole.
The time $t$ is the  Rindler boost-angle time. The corresponding energy is also dimensionless and is equal to the entropy, which in the qubit model is equal to the number of qubits $K.$

For quantum circuits with parallel Hayden-Preskill architecture, $t$ also has special significance. In that case $t$ has the significance of the clock time which ticks off one unit for every step in which there are of order $K$ gates. 
 For sub-exponential times  the rate of complexity growth in these units is $\sim K.$

\bn
10) The circuit complexity is denoted by $\CC.$  The Kolmogorov complexity of a string $s$ is denoted 
$\CC_{\kappa}(s).$

\bn
11) $B_a$ is the coefficient in the Gaussian probability distribution of the coupling constants $J$. It is also the inverse temperature of the $\CA$-model. $T_a=1/B_a$ is the temperature of the $\CA$-model.

\bn
12)
$\CI_{IJ} $ is a symmetric matrix in the adjoint representation of 
\sk.  It is called the moment of inertia tensor.

\bn
13)
Gates are denoted $g.$ A sequence of $n$ gates forming a circuit is denoted $ g_n g_{n-1}....g_1 $. 

\bn
14)
$e^{\Lambda}$ denotes a Loschmidt-echo operator defined by  $e^{\Lambda}=e^{-iHt}e^{i(H+\Delta) t}$.

\bn
15) By a Hayden-Preskill circuit we mean a circuit of $K$ qubits such that in each time-step the qubits are paired and interact through $K/2$ gates \cite{Hayden:2007cs}. (This is the version with 2-local gates; it can be generalized to a version with $k$-local gates in which at each time-step the qubits are sorted into groups of $k$ and interact through $K/k$ gates.)

\bn
16) The `uncomplexity' is the amount $\Delta \CC$ by which the computational complexity $\CC$ of a state is less than the maximum possible complexity for that state, as in Eq.~\ref{eq:definitionofuncomplexity}
\begin{equation}
\Delta \CC =  \CC_\textrm{max} - \CC.
\end{equation}

\sc
\section{Some Clarifications}\label{Sec:clarifications}
After we initially circulated this paper some questions came up from colleagues that we find worth discussing.

 \begin{enumerate}

\item
 The first concerns Eq.~\ref{eq:$k$-local} and the definition of \kl. The expression in Eq.~\ref{eq:$k$-local} contains only terms of weight $k$ whereas the standard definition of \kl \ allows all terms of weight up to and including $k.$ In several places throughout the paper the equations refer to the more restricted version of `exact' $k$ locality---only operators of weight $k$ in the Hamiltonian---but they can be easily generalized to accommodate the more general case.
\item
 The choice of time units that we use throughout is motivated by black hole physics, where the dimensionless Rindler time $t$ is defined in terms of the Schwarzschild time $t_{schw}$ by
\be 
t=\frac{t_{schw}}{2\pi \beta},
\ee
as in Eq.~\ref{eq:tintermsoftschw}; here $\beta$ is the inverse Hawking temperature of the black hole. We may also think of $t$ as measuring the number of time steps in a Hayden-Preskill circuit.  With such a choice of units the Hamiltonian and the $J$ coefficients are also dimensionless.
\item 
We have been asked why the draconian choice of penalty factors in \cite{Dowling} is inconsistent with the switchback effect. To understand this we remind the reader how the circuit  complexity of precursors
evolves for times earlier than the scrambling time (for a review see \cite{Brown:2016wib}). The complexity of precursors grows very slowly until the scrambling time, and then suddenly begins to increase linearly. (The same is true for Loschmidt echo operators.)
Before the scrambling time the growth rate is not zero but is negligible. However, draconian penalty factors of order $4^K$ would punish shortcuts exponentially harshly. With shortcuts effectively forbidden, the complexity would grow linearly almost immediately. In order to agree with the complexity growth for  discrete quantum circuits we need the penalty factors in the continuous Hamiltonian theory to turn on much more smoothly. We will come back to this point in \cite{UsInFuture}.
\item
Another question that came up is why, in Sec.~\ref{one clean qubit},  we do not consider the measurement at the end of a computation as part of computational work. For example why do we not allow a complete measurement that  re-initializes the computer to a random simple state? The answer is that measurement is not something that is part of the auxiliary description of the quantum evolution. But more important, from the global point of view the measurement is equivalent to the development of entanglement of $\CQ$ with the rest of the world. To follow the resource we would have to consider the changes in the complexity of everything, including the observer. We believe that if we did so the overall complexity would increase when a measurement is done, and that would cause a decrease in the global version of the resource.
\item
The circuit depicted in Fig.~\ref{C1} does not obviously look like a \kl \ circuit. However let us define the $(k+1)$-local gate 
\begin{equation}
\tilde{g_n} = H g_{nc} H,
\end{equation}
where $H$ is the Hadamard gate and $g_{nc}$ is the \kl \ gate $g_n$ controlled by $\tau.$ Then if the operator $G$ is a product of \kl \ gates
$$
G=...g_5 g_4 g_3 g_2 g_1,
$$
the circuit in Fig.~\ref{C1} is equivalent to the $(k+1)$-local circuit 
\be 
...\tilde{g_5}\tilde{g_4}\tilde{g_3}\tilde{g_2}\tilde{g_1}.
\label{tilde_ggg}
\ee
We may therefore think of the computational work as being done in small $(k+1)$-local  steps, each using a small amount of the resource.

It is interesting to view the computation from the point of view of the 2-gas model in Fig.~\ref{f4}. The effect of the operations in Eq.~\ref{tilde_ggg} is to evolve one of the component gases according to the circuit 
$...g_5 g_4 g_3 g_2 g_1$ while leaving the other component fixed. Since the initial state $|\rm max\ra$ is maximally complex the fixed gas is already in equilibrium. The effect of the circuit Eq.~\ref{tilde_ggg} is to break the correlation between the two components, and if it goes on long enough, to bring the whole system to equilibrium.

\end{enumerate}

\sc
\section{Action vs. Distance in the Toy Model}\label{sec: A vs D}

(Note about conventions: In \cite{Brown:2016wib} the time variable was called $\tau$ while in this paper the same variable is called $t$.)

In the original version of complexity geometry, complexity was identified with geodesic distance from the identity. In \cite{Brown:2016wib} we remarked that there is an alternative formulation in which complexity is identified with the action along a geodesic. However the analysis was carried out with the earlier formulation. Since  this paper uses the action formulation, there is a minor difference of conventions between   \cite{Brown:2016wib} and the present paper. The difference between the two formulations can be absorbed into a re-definition of the scale of the metric. 

In the toy model the complexity geometry is simplified to a two-dimensional geometry with uniform negative curvature. The metric has  the form
\be 
dl^2 = F^2 (dr^2 + \sinh^2{r} \  d\theta^2).
\ee

Consider two neighboring geodesics passing through $r=0.$ The distance between them grows like $e^r$ which can be written as 
\be 
d(t) = e^{\dot{r}t}.
\ee 
This identifies $\dot{r}$ as the Lyapunov exponent controlling scrambling. With our choice of dimensionless time (Rindler time ) the Lyapunov exponent is $1$ which constrains the motion to satisfy
\be 
\dot{r} =1.
\ee

The two formulations can be expressed as follows:

\bn
Distance formulation:
\be 
\CC = F\int \sqrt{(\dot{r}^2 + \dot{\theta}^2 \ \sinh^2{r}) \ } dt.
\label{distance-form}
\ee

\bn
Action formulation:
\be 
\CC = \frac{1}{2}F^2\int {(\dot{r}^2 + \dot{\theta}^2 \ \sinh^2{r}) \ }dt.
\label{action-form}
\ee

In both cases the complexity should grow like $Kt.$ In the distance formulation this requires $F=K.$ This is equivalent to the curvature being $\sim -1/K^2.$

In the action formulation the growth of complexity requires 
\be 
\frac{F^2}{2} = K
\ee
and the curvature is $\sim -1/K$ in agreement with the calculation in 
Sec.~\ref{Sec: sectional curvature}.

There is no inconsistency between the two formulations. The difference can be absorbed into the normalization of the metric Eq.~\ref{bimetric}. If we wish to use distance rather than action we would need to modify Eq.~\ref{bimetric} to
$$
dl^2 = K \Tr [d\Ud dU].
$$
Such a change would have no effect on the agreement between the curvature and the calculation in Sec.~\ref{Sec: sectional curvature}.


\end{document}